\documentclass[twocolumn]{aastex631}
\usepackage{amsmath,amssymb}
\usepackage{graphicx}

\shorttitle{Unusual $\alpha^2$ CVn Stars}
\shortauthors{Heinze et al.}

\begin{document}

\title{Most Rotational Variables Dominated by a Single Bright Feature are $\alpha^2$ CVn Stars}

\author{A. N. Heinze}
\affiliation{DiRAC Institute and the Department of Astronomy, University of Washington, 3910 15th Avenue NE, Seattle, WA 98195; aheinze@uw.edu}

\author{Heather Flewelling}
\affiliation{Canada-France-Hawaii Telescope, 65-1238 Mamalahoa Hwy, Kamuela, HI 96743}

\author{Mark E. Huber}
\affiliation{Institute for Astronomy, University of Hawaii, 2680 Woodlawn, Honolulu, HI, 96822, USA}

\correspondingauthor{A. N. Heinze} 

\begin{abstract}
We previously reported a rare class of variable star light curves isolated from a sample of 4.7 million candidate variables from the ATLAS survey. Dubbed `UCBH' light curves, they have broad minima and narrow, symmetrical maxima, with typical periods of 1-10 days and amplitudes of 0.05--0.20 mag. They maintain constant amplitude, shape, and phase coherence over multiple years, but do not match any known class of pulsating variables. A localized bright spot near the equator of a rotating star will produce a UCBH-type light curve for most viewing geometries. Most stars that exhibit rotational variability caused primarily by a single bright feature should therefore appear as UCBH stars, although a rotating bright spot is not the only thing that could produce a UCBH-type lightcurve. We have spectroscopically investigated fourteen UCBH stars and found ten of them to be Ap/Bp stars: A-type or B-type stars with greatly enhanced photospheric abundances of specific heavy elements. Rotationally variable Ap/Bp stars are referred to as $\alpha^2$ CVn variables. Most ATLAS UCBH stars are therefore $\alpha^2$ CVn stars, although only a minority of $\alpha^2$ CVn stars in the literature have UCBH light curves. The fact that $\alpha^2$ CVn stars dominate the UCBH class suggests that lone bright spots with sufficient size and contrast develop more readily on Ap/Bp stars than on any other type. The $\alpha^2$ CVn UCBH stars may be characterized by a specific magnetic field topology, making them intriguing targets for future Zeeman-Doppler imaging. \end{abstract}


\section{Introduction}

Though stellar photometry is typically not their primary mission, modern astronomical surveys such as the Catalina Sky Survey \citep{CSS}, the All-Sky Automated Survey for Supernovae \citep[ASAS-SN,][]{Shappee2014}, Pan-STARRS1 \citep{PS1A,Flewelling2016,PS1B,PS1C,PS1D}, ATLAS \citep{Tonry2018}, the Zwicky Transient Facility \citep{ZTF}, and others produce well-sampled photometric time series for millions of stars. These data sets are invaluable both for large-scale statistics of variables stars and for identifying rare, highly interesting objects. The huge sample sizes and the presence of photometry but not spectra for many of the objects enable an interesting new perspective on variable stars. Spectrum-blind analysis of millions of light curves can reveal new, physically meaningful commonalities that do not necessarily align with established classes of variable stars. Though the established classes are (of course) also physically meaningful, they were defined in a context of smaller sample sizes and more intensive spectroscopic investigation to which the current big-data context has meaningful things to add.

Herein, we analyze a rare class of variable stars, the `UCBH' stars \citep{ATLASvar}, defined by a specific lightcurve shape and identified purely photometrically using data from the ATLAS survey. We introduce these stars in Section \ref{sec:UCBHintro}, and in Section \ref{sec:ACVNintro} we introduce the established variable class (the $\alpha^2$ CVn stars) to which most of them are found to belong. In Section \ref{sec:char} we show examples of UCBH light curves and demonstrate that the characteristic light curve shape will result from a single bright spot on a rotating star, over a wide range of sizes and viewing geometries. We present our spectroscopic results in Section \ref{sec:spec}, demonstrating that most of them are $\alpha^2$ CVn stars (although only a minority of known $\alpha^2$ CVn stars have UCBH-type lightcurves). In Section \ref{sec:HR} we use Gaia parallaxes to place our UCBH stars on HR diagrams, demonstrating that most of them have luminosities and colors consistent with main-sequence Ap stars subject to interstellar reddening --- with some interesting exceptions. We discuss astrophysical implications and offer our conclusions in Section \ref{sec:conc}.

\begin{figure*} 
\plottwo{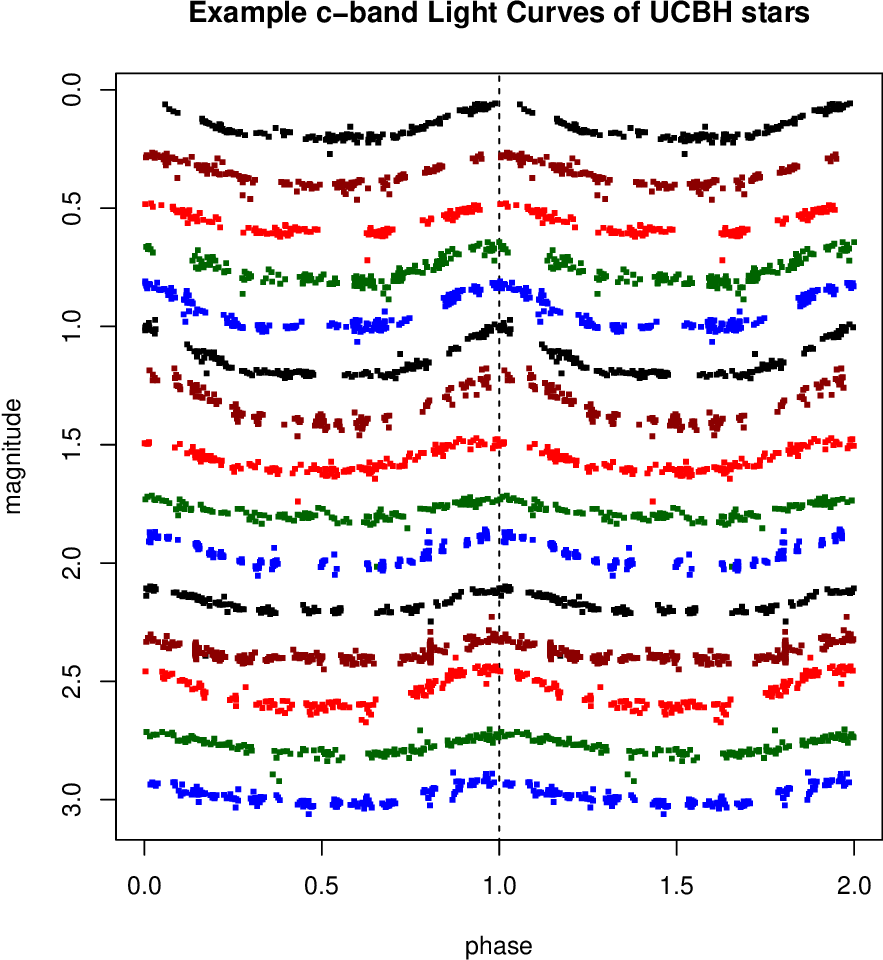}{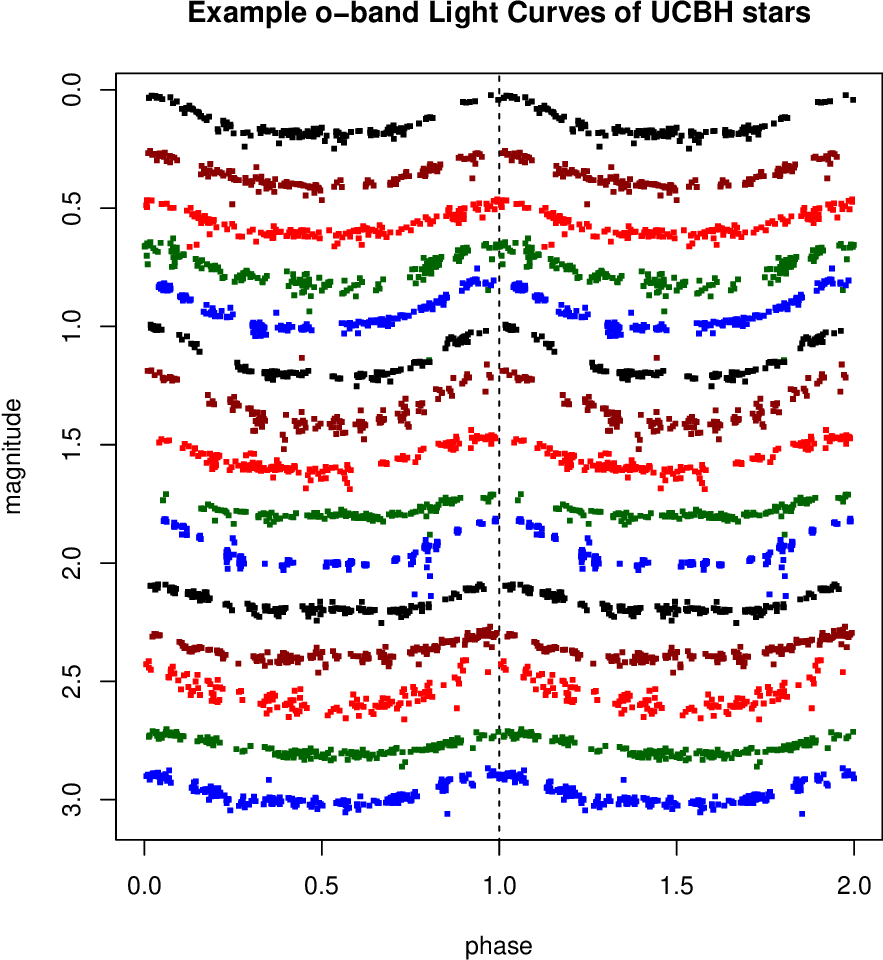}
\caption{Example lightcurves for ATLAS UCBH stars. The left panel shows the c-band lightcurves, and the right panel shows the corresponding o-band lightcurves for the same objects. A random selection of stars has been attempted to avoid cherry-picking the cleanest examples.}
\label{fig:many} 
\end{figure*}

\subsection{ATLAS Variable Star DR1 and the UCBH stars} \label{sec:UCBHintro}

The Asteroid Terrestrial-impact Last Alert System \cite[ATLAS;][]{Tonry2018} is a NASA-funded planetary defense survey that scans the sky for near-Earth asteroids while simultaneously producing well-calibrated data useful for many other astrophysical investigations. Each ATLAS image is photometrically calibrated using a customized, highly-precise catalog \citep{refcat} created by mutually calibrating several state-of-the-art photometric catalogs. 

In its first two years on the sky, ATLAS operated only one telescope (it now has four). This single ATLAS unit surveyed one fourth of the accessible sky every night, obtaining four 30-second exposures of each target field over a period of about one hour. Hence, during good weather in its observing season, a given star would be observed an average of once per night -- but these observations occur in clumps of four in one hour, with a four-day gap before the next clump. Not all images yield flux measurements of every object in the field: for example, faint stars would not be detected in bad seeing. Nevertheless, in two years ATLAS obtained 100 or more photometric measurements for each of 142 million distinct stars, of which 4.7 million were identified as candidate variables. Photometric time series for these candidate variables, as well as classifications we obtained for them using machine learning, constitute ATLAS variable star Data Release One (DR1) and are publicly available through STScI \citep{ATLASvar}.

While preparing ATLAS DR1, we manually examined thousands of light curves of objects that had periods, amplitudes, or other characteristics not typical of the classes the machine had assigned them. We identified a rare but well-defined class of light curves, mostly identified as pulsators by the machine, that did not seem to match any known type of variable star. These objects had coherent, periodic light curves with a distinctive shape defined by narrow, symmetrical maxima and broad, flat minima (Figure \ref{fig:many}). They looked like the light curves of contact eclipsing binaries turned upside down. Since we'd defined a light curve category called CBH ({\underline C}ontact eclipsing {\underline B}inaries folded at {\underline H}alf the true period), we called this new set of stars the upside-down CBH variables, or UCBH stars. They have typical periods of 1--10 days and peak-to-trough amplitudes of 0.05--0.20 magnitudes. The amplitudes are usually similar between the ATLAS $c$ and $o$ bands\footnote{These broad, customized bandpasses are described in \citet{Tonry2018}; briefly, $c$ corresponds approximately to Sloan $g+r$ and $o$ to $r+i$.}.

Herein, we present a catalog of 98 UCBH stars identified in ATLAS DR1 photometry. This catalog constitutes the entire set of ATLAS variables we have confidently assigned to the UCBH class. We carry the analysis of UCBH stars beyond pure photometry for the first time, presenting low-resolution spectra for 14 of them (chosen based on brightness and observability during our scheduled telescope time), intensive multi-band photometry for one, and HR diagrams based on Gaia parallaxes for all. 

\subsection{Overview of $\alpha^2$ CVn Variables} \label{sec:ACVNintro}

Our spectra (Section \ref{sec:spec}) indicate that a majority of UCBH stars are $\alpha^2$ CVn variables. An $\alpha^2$ CVn variable is an Ap or Bp star that exhibits rotationally modulated variability \citep{Peterson1970,Catalano1993}. Ap and Bp stars are A-type or B-type stars with enormously enhanced photospheric abundances of specific heavy elements (silicon, chromium, strontium, europium, and others). The enhancement is believed to be produced by radiative levitation \citep{Michaud1970}. This levitation occurs because the elements in question interact more strongly with the radiation field than most other atoms --- i.e., they have many strong spectral lines at wavelengths near the peak of the star's spectral energy distribution \citep{Hummerich2018}. Radiation pressure therefore exerts a stronger upward force (relative to their mass) on the atoms of these elements than on the majority constituents of the stellar atmosphere.  This upward force is believed to concentrate the elements in the upper layers of the stars. The stellar atmospheres must be remarkably free from convection for the extremely weak force of radiative levitation to produce the observed concentrations of heavy elements. \citet{Michaud1970} calculated that convective velocities must be slower than $10^{-5}$ m/sec, and theorized that strong magnetic fields might be able to stabilize the ionized atmospheres of these stars against convective stirring.

That some Ap/Bp stars should exhibit rotational variability is not surprising: longitudinal inhomogeneity in the concentration of heavy elements would naturally cause rotational variation \citep{Peterson1970}. The shape, amplitude, and detectability of this variation depend on the details of the inhomogeneity \citep{Shulyak2010}, which in turn might arise from spatial variations in the magnetic field \citep{Michaud1981}.

Interestingly, \citet{Peterson1970} found that a photospheric spot having an enhanced concentration of silicon will be {\em brighter} at optical wavelengths, because the strong silicon absorption lines in the UV will redistribute flux into the optical. We might expect that the heavy elements would be most concentrated at the regions of strongest magnetic field (where convection is most strongly suppressed and radiative levitation can have the greatest effect), and that the flux redistribution would render these regions the brightest at optical wavelengths. This would imply that the (optical) photometric maximum should coincide with points when the region of greatest magnetic field strength is centered on the hemisphere of the star that faces us. Accordingly, \citet{Dukes2018} found that the $\alpha^2$ CVn star HD 215441 has its photometric maximum at about the same rotational phase as the maximum of the magnetic field measured by Zeeman splitting of the spectral lines. This might not, however, be a general rule: both theory \citep{Michaud1981} and observation \citep{Kochukhov2010,Kochukhov2015} indicate that different radiatively levitated elements can be affected differently by the magnetic field and hence have different photospheric distributions. Despite this complexity, there is broad observational evidence \citep[e.g.][]{Pyper1969} that optical photometric maxima do tend to occur near the same rotational phase where the greatest abundance of radiatively levitated elements is measured, consistent with the UV flux redistribution predicted by \citet{Peterson1970}.

Known $\alpha^2$ CVn stars have amplitudes mostly smaller than is typical for the ATLAS UCBH stars, and some have longer periods, but the distributions of both period and amplitude overlap heavily. \citet{Sikora2019} have shown that the rotation periods of most of the magnetic, chemically peculiar A and B stars whose variability enables their periods to be measured fall within the same 1--10 day range that characterizes UCBH stars, with the few exceptions mostly having longer periods.

The light curves of known $\alpha^2$ CVn stars have a variety of shapes, including some that exactly match our UCBH stars and many that do not. \citet{Hensberge1977} present $uvby$ photometry of six $\alpha^2$ CVn stars with periods from 1.48 to 4.75 days, of which only one (HD 207188) shows a UCBH-type lightcurve. \citet{Ryabchikova1990} find the $\alpha^2$ CVn star HD 192913 to have a period of 16.5 days and an amplitude of 0.04 magnitudes, with a saw-tooth rather than UCBH-type light curve. \citet{Catalano1993} present multi-band ($uvby$) light curves of eight bright $\alpha^2$ CVn stars with periods ranging from 1.3 to 6.8 days and amplitudes from 0.03--0.10 mag. Two of them (HD 54118 and HD 73340) exhibit UCBH-type light curves in at least one of the four photometric bands, while the others have various different shapes. \citet{Poretti1997} also probe $\alpha^2$ CVn stars with $uvby$ photometry: HR 2746 (period 0.92 days; amplitude 0.004--0.024 mag depending on photometric band) and HR 2761 (period 2.06 days; amplitudes 0.015--0.057 mag). They find a UCBH-type lightcurve for HR 2746 and sinusoidal variations for HR 2761. \citet{Drury2017} present Kepler and ground-based photometry of the $\alpha^2$ CVn star KIC 2569073, finding a period of 14.67 days, approximately sinusoidal variations, and peak-to-trough amplitudes varying from 0.03 to 0.34 magnitudes, with a phase reversal seen in the $B$-band relative to the $V$, $R_{\mathrm{C}}$, and $I_{\mathrm{C}}$ bands. \citet{Dukes2018} acquired precise $uvby$ lightcurves of eight $\alpha^2$ CVn stars. One of these, HD 26792, shows a perfect UCBH-type light curve in all filters, while HD 5797 shows a noisy but UCBH-like light curve in the $y$ filter only. Among the other six stars, none shows a UCBH-type light curve shape in any filter. Most recently, \citet{Bernhard2020} analyzed and published light curves for 294 magnetic chemically peculiar stars (i.e., Ap/Bp stars) using data from three recent surveys that used very small apertures and hence maintained photometric precision for very bright stars. We find that 33 of the \citet{Bernhard2020} light curves are of the UCBH type (more on this in Section \ref{sec:bernhard}). Finally, after this work was accepted for publication, we learned of another paper \citep{Bernhard2021} demonstrating that many of the ATLAS UCBH stars are $\alpha^2$ CVn variables. While this same result is also an important conclusion of the current work, our very different emphasis makes our work and that of \citet{Bernhard2021} substantially complementary.

\section{The Characteristic Light Curves of UCBH Stars} \label{sec:char}

In Tables \ref{tab:yescat} and \ref{tab:nocat} we present the full set of UCBH stars we have identified in ATLAS data, divided into those that are (Table \ref{tab:yescat}) and are not (Table \ref{tab:nocat}) probable $\alpha^2$ CVn stars, based on color and luminosity thresholds discussed in Section \ref{sec:HR}.

\startlongtable
\begin{deluxetable*}{llcccccccc}
\tablewidth{0pt}
\tabletypesize{\small}
\tablecaption{ATLAS UCBH stars likely to be $\alpha^2$ CVn stars\label{tab:yescat}}
\tablehead{\colhead{ATLAS ID\tablenotemark{a}} & \colhead{Period (d)} & \colhead{$c-o$\tablenotemark{b}} & \colhead{amplitude\tablenotemark{c}} & \colhead{g} & \colhead{r} & \colhead{g-z} & \colhead{parallax\tablenotemark{d} (mas)} & \colhead{$M_V$\tablenotemark{e}}& \colhead{$M_K$\tablenotemark{e}}}
\startdata
J010.7230+57.8087 & 4.435267 & -0.004 & 0.208 & 12.962 & 12.831 & -0.03 & $0.544 \pm 0.014$ & $1.56_{-0.05}^{+0.05}$ & $1.01_{-0.05}^{+0.05}$ \\
J042.3381+51.3632 & 2.762592 & 0.069 & 0.172 & 13.762 & 13.581 & 0.08 & $0.389 \pm 0.018$ & $1.60_{-0.11}^{+0.10}$ & $0.88_{-0.11}^{+0.10}$ \\
J053.4996+56.7983 & 3.709273 & 0.372 & 0.176 & 12.516 & 12.113 & 0.74 & $0.726 \pm 0.141$ & $1.58_{-0.47}^{+0.39}$ & $-0.09_{-0.47}^{+0.39}$ \\
J060.4363+55.5067 & 2.474011 & 0.329 & 0.126 & 14.202 & 13.778 & 0.78 & $0.527 \pm 0.015$ & $2.56_{-0.06}^{+0.06}$ & $0.87_{-0.06}^{+0.06}$ \\
J061.6000+59.6651 & 3.434188 & 0.158 & 0.191 & 14.400 & 14.097 & 0.35 & $0.274 \pm 0.023$ & $1.41_{-0.19}^{+0.18}$ & $0.22_{-0.19}^{+0.18}$ \\
J062.2757+57.3439 & 2.119598 & 0.281 & 0.131 & 14.863 & 14.497 & 0.54 & $0.262 \pm 0.020$ & $1.74_{-0.17}^{+0.16}$ & $0.39_{-0.17}^{+0.16}$ \\
J063.5805+46.9075 & 1.456892 & 0.271 & 0.145 & 13.601 & 13.237 & 0.58 & $0.274 \pm 0.082$ & $0.58_{-0.77}^{+0.57}$ & $-0.88_{-0.77}^{+0.57}$ \\
J065.5257+51.2992 & 3.622927 & 0.358 & 0.209 & 15.384 & 14.837 & 0.77 & $0.324 \pm 0.024$ & $2.62_{-0.17}^{+0.16}$ & $0.78_{-0.17}^{+0.16}$ \\
J065.7038+47.6938 & 2.774777 & 0.174 & 0.154 & 11.687 & 11.452 & 0.34 & $1.222 \pm 0.189$ & $1.98_{-0.37}^{+0.31}$ & $0.79_{-0.37}^{+0.31}$ \\
J065.8718+43.5268 & 1.453359 & 0.157 & 0.115 & 14.617 & 14.437 & 0.23 & $0.269 \pm 0.021$ & $1.66_{-0.18}^{+0.16}$ & $0.63_{-0.18}^{+0.16}$ \\
J072.5642+39.5294\tablenotemark{f} & 2.263109 & 0.220 & 0.143 & 14.852 & 14.597 & 0.43 & $-0.022 \pm 0.123$ & $<2.41$ & $<1.10$ \\
J073.7460+43.3008 & 7.746157 & 0.011 & 0.165 & 13.697 & 13.624 & -0.04 & $0.322 \pm 0.018$ & $1.19_{-0.12}^{+0.12}$ & $0.54_{-0.12}^{+0.12}$ \\
J076.4023+45.6101 & 2.156734 & 0.222 & 0.158 & 14.793 & 14.528 & 0.38 & $0.322 \pm 0.024$ & $2.18_{-0.17}^{+0.16}$ & $0.83_{-0.17}^{+0.16}$ \\
J079.6501+37.6469 & 2.445079 & 0.252 & 0.076 & 14.636 & 14.289 & 0.56 & $0.290 \pm 0.037$ & $1.74_{-0.29}^{+0.26}$ & $0.18_{-0.29}^{+0.26}$ \\
J080.3836+43.4165 & 2.896787 & 0.112 & 0.073 & 14.607 & 14.443 & 0.19 & $0.174 \pm 0.023$ & $0.71_{-0.31}^{+0.27}$ & $-0.28_{-0.31}^{+0.27}$ \\
J081.9629+42.4325 & 1.761931 & 0.024 & 0.187 & 13.104 & 13.071 & -0.02 & $0.537 \pm 0.021$ & $1.73_{-0.08}^{+0.08}$ & $1.06_{-0.08}^{+0.08}$ \\
J082.4358+39.0290 & 1.947387 & 0.313 & 0.169 & 15.162 & 14.827 & 0.64 & $0.350 \pm 0.025$ & $2.68_{-0.16}^{+0.15}$ & $0.95_{-0.16}^{+0.15}$ \\
J084.6836+33.5786 & 1.083538 & 0.159 & 0.072 & 14.163 & 13.938 & 0.31 & $0.422 \pm 0.019$ & $2.16_{-0.10}^{+0.09}$ & $0.94_{-0.10}^{+0.09}$ \\
J084.9321+37.2119 & 1.838578 & 0.348 & 0.203 & 15.146 & 14.674 & 0.79 & $0.441 \pm 0.027$ & $3.09_{-0.14}^{+0.13}$ & $1.17_{-0.14}^{+0.13}$ \\
J089.1590+29.6893 & 2.092882 & 0.111 & 0.201 & 15.228 & 15.023 & 0.37 & $0.221 \pm 0.029$ & $1.83_{-0.31}^{+0.27}$ & $0.73_{-0.31}^{+0.27}$ \\
J089.1800+11.3598 & 6.858919 & -0.083 & 0.200 & 12.974 & 12.993 & -0.22 & $0.438 \pm 0.021$ & $1.19_{-0.11}^{+0.10}$ & $0.83_{-0.11}^{+0.10}$ \\
J089.7490+22.3852 & 1.828602 & 0.216 & 0.067 & 14.811 & 14.530 & 0.47 & $0.322 \pm 0.021$ & $2.19_{-0.15}^{+0.14}$ & $0.87_{-0.15}^{+0.14}$ \\
J089.7821+14.0001 & 2.914186 & -0.025 & 0.167 & 14.139 & 14.109 & -0.13 & $0.241 \pm 0.018$ & $1.02_{-0.17}^{+0.16}$ & $0.52_{-0.17}^{+0.16}$ \\
J090.0237+21.9017 & 1.852552 & 0.344 & 0.081 & 15.117 & 14.644 & 0.80 & $0.415 \pm 0.026$ & $2.93_{-0.14}^{+0.13}$ & $1.12_{-0.14}^{+0.13}$ \\
J090.7799+32.5180 & 3.392613 & 0.190 & 0.204 & 15.928 & 15.612 & 0.48 & $0.205 \pm 0.039$ & $2.30_{-0.45}^{+0.38}$ & $1.02_{-0.45}^{+0.38}$ \\
J091.4130+26.0389 & 1.653195 & 0.048 & 0.070 & 14.053 & 13.973 & 0.04 & $0.326 \pm 0.026$ & $1.57_{-0.18}^{+0.17}$ & $0.75_{-0.18}^{+0.17}$ \\
J091.6776+11.6763 & 2.546904 & 0.131 & 0.222 & 15.743 & 15.453 & 0.33 & $0.228 \pm 0.032$ & $2.36_{-0.33}^{+0.28}$ & $1.47_{-0.33}^{+0.28}$ \\
J092.1616+30.8849 & 2.167543 & 0.288 & 0.130 & 14.564 & 14.178 & 0.60 & $0.328 \pm 0.022$ & $1.92_{-0.15}^{+0.14}$ & $0.40_{-0.15}^{+0.14}$ \\
J095.0108+01.8715 & 2.379236 & 0.179 & 0.164 & 14.074 & 13.783 & 0.45 & $0.374 \pm 0.017$ & $1.77_{-0.10}^{+0.10}$ & $0.52_{-0.10}^{+0.10}$ \\
J095.1623+09.4416 & 2.680946 & -0.013 & 0.055 & 13.176 & 13.154 & -0.06 & $0.252 \pm 0.016$ & $0.17_{-0.14}^{+0.14}$ & $-0.52_{-0.14}^{+0.14}$ \\
J098.3181-07.1966 & 1.777259 & 0.163 & 0.084 & 12.823 & 12.596 & 0.29 & $0.736 \pm 0.014$ & $2.02_{-0.04}^{+0.04}$ & $0.92_{-0.04}^{+0.04}$ \\
J098.5294-00.6734 & 1.756423 & 0.073 & 0.172 & 12.441 & 12.305 & 0.07 & $0.721 \pm 0.014$ & $1.65_{-0.04}^{+0.04}$ & $0.91_{-0.04}^{+0.04}$ \\
J098.8005+06.4102 & 5.645326 & 0.179 & 0.134 & 13.963 & 13.719 & 0.35 & $0.390 \pm 0.014$ & $1.77_{-0.08}^{+0.08}$ & $0.43_{-0.08}^{+0.08}$ \\
J099.2222+00.8030 & 8.34857 & 0.273 & 0.153 & 13.815 & 13.441 & 0.57 & $0.454 \pm 0.022$ & $1.88_{-0.11}^{+0.10}$ & $0.37_{-0.11}^{+0.10}$ \\
J100.1756+00.2310 & 1.461506 & 0.141 & 0.075 & 14.051 & 13.857 & 0.28 & $0.428 \pm 0.020$ & $2.09_{-0.10}^{+0.10}$ & $0.97_{-0.10}^{+0.10}$ \\
J105.5347-01.9077 & 2.944283 & -0.035 & 0.102 & 13.902 & 13.903 & -0.17 & $0.260 \pm 0.015$ & $0.97_{-0.13}^{+0.13}$ & $0.47_{-0.13}^{+0.13}$ \\
J106.0648-03.3054 & 1.428092 & 0.075 & 0.120 & 15.109 & 14.961 & 0.14 & $0.216 \pm 0.033$ & $1.69_{-0.37}^{+0.31}$ & $0.80_{-0.37}^{+0.31}$ \\
J106.2127-00.9740 & 2.027484 & -0.062 & 0.129 & 13.672 & 13.707 & -0.21 & $0.404 \pm 0.016$ & $1.72_{-0.09}^{+0.09}$ & $1.27_{-0.09}^{+0.09}$ \\
J107.2480-12.7673 & 2.951689 & 0.298 & 0.096 & 14.592 & 14.194 & 0.67 & $0.320 \pm 0.018$ & $1.88_{-0.13}^{+0.12}$ & $0.13_{-0.13}^{+0.12}$ \\
J108.1707-08.2617 & 2.079303 & 0.105 & 0.079 & 14.238 & 14.074 & 0.22 & $0.288 \pm 0.020$ & $1.44_{-0.16}^{+0.15}$ & $0.28_{-0.16}^{+0.15}$ \\
J109.4314-15.7080 & 3.587642 & 0.311 & 0.137 & 15.395 & 14.981 & 0.71 & $0.234 \pm 0.027$ & $2.00_{-0.27}^{+0.24}$ & $0.27_{-0.27}^{+0.24}$ \\
J109.7734-07.1470 & 1.858906 & -0.096 & 0.059 & 13.978 & 14.049 & -0.28 & $0.255 \pm 0.021$ & $1.05_{-0.19}^{+0.17}$ & $0.76_{-0.19}^{+0.17}$ \\
J110.2675-03.2520 & 2.809982 & -0.146 & 0.233 & 12.659 & 12.753 & -0.39 & $0.445 \pm 0.024$ & $0.95_{-0.12}^{+0.11}$ & $0.71_{-0.12}^{+0.11}$ \\
J110.2057-08.5343 & 2.501694 & 0.031 & 0.167 & 15.241 & 15.003 & 0.13 & $0.284 \pm 0.027$ & $2.36_{-0.22}^{+0.20}$ & $1.58_{-0.22}^{+0.20}$ \\
J110.9074-12.0800 & 1.647087 & 0.055 & 0.075 & 13.760 & 13.655 & 0.10 & $0.347 \pm 0.018$ & $1.39_{-0.11}^{+0.11}$ & $0.54_{-0.11}^{+0.11}$ \\
J110.9392-21.9535 & 1.609184 & 0.229 & 0.084 & 14.555 & 14.269 & 0.43 & $0.333 \pm 0.018$ & $2.00_{-0.12}^{+0.11}$ & $0.55_{-0.12}^{+0.11}$ \\
J111.2165-23.1396 & 7.770775 & 0.311 & 0.120 & 15.032 & 14.636 & 0.66 & $0.215 \pm 0.018$ & $1.46_{-0.19}^{+0.18}$ & $-0.25_{-0.19}^{+0.18}$ \\
J112.7672-19.4071 & 1.769319 & 0.065 & 0.098 & 13.521 & 13.425 & 0.06 & $0.364 \pm 0.012$ & $1.27_{-0.08}^{+0.07}$ & $0.33_{-0.08}^{+0.07}$ \\
J112.9447-28.3476\tablenotemark{f} & 2.963217 & -0.008 & 0.110 & 14.224 & 14.185 & -0.08 & $-0.272 \pm 0.181$ & $<1.36$ & $<0.57$ \\
J113.2387-19.4340 & 2.273453 & 0.150 & 0.082 & 13.536 & 13.309 & 0.26 & $0.353 \pm 0.013$ & $1.14_{-0.08}^{+0.08}$ & $0.06_{-0.08}^{+0.08}$ \\
J113.8870-24.3497 & 1.424931 & 0.145 & 0.136 & 14.903 & 14.674 & 0.31 & $0.251 \pm 0.019$ & $1.76_{-0.17}^{+0.16}$ & $0.60_{-0.17}^{+0.16}$ \\
J114.4848-21.4294 & 2.972624 & 0.043 & 0.110 & 14.070 & 13.953 & 0.06 & $0.290 \pm 0.014$ & $1.31_{-0.11}^{+0.10}$ & $0.54_{-0.11}^{+0.10}$ \\
J115.0928-10.6198 & 1.908329 & -0.166 & 0.083 & 14.320 & 14.439 & -0.47 & $0.269 \pm 0.024$ & $1.54_{-0.20}^{+0.18}$ & $1.34_{-0.20}^{+0.18}$ \\
J116.4925-16.0440 & 1.807299 & -0.120 & 0.099 & 14.584 & 14.684 & -0.28 & $0.237 \pm 0.024$ & $1.51_{-0.23}^{+0.21}$ & $1.21_{-0.23}^{+0.21}$ \\
J116.5342-19.8451\tablenotemark{g} & 1.860305 & 0.082 & 0.075 & 13.969 & 13.825 & 0.13 & \nodata & \nodata & \nodata \\
J116.5376-20.4285 & 1.682906 & 0.016 & 0.137 & 14.029 & 13.954 & -0.05 & $0.342 \pm 0.016$ & $1.65_{-0.10}^{+0.10}$ & $0.96_{-0.10}^{+0.10}$ \\
J116.8586-29.7166 & 2.245591 & 0.314 & 0.117 & 14.672 & 14.223 & 0.68 & $0.293 \pm 0.016$ & $1.75_{-0.12}^{+0.11}$ & $0.03_{-0.12}^{+0.11}$ \\
J117.0754-24.7339 & 1.521412 & -0.007 & 0.089 & 13.216 & 13.159 & -0.10 & $0.242 \pm 0.020$ & $0.10_{-0.19}^{+0.17}$ & $-0.58_{-0.19}^{+0.17}$ \\
J117.1588-19.3073 & 2.776477 & 0.014 & 0.115 & 14.308 & 14.202 & -0.02 & $0.264 \pm 0.019$ & $1.35_{-0.16}^{+0.15}$ & $0.66_{-0.16}^{+0.15}$ \\
J117.4249-26.8279 & 2.784376 & 0.049 & 0.217 & 14.771 & 14.536 & 0.16 & $0.226 \pm 0.017$ & $1.40_{-0.17}^{+0.16}$ & $0.59_{-0.17}^{+0.16}$ \\
J117.8237-09.5101 & 1.789376 & -0.203 & 0.117 & 12.881 & 13.062 & -0.53 & $0.496 \pm 0.033$ & $1.46_{-0.15}^{+0.14}$ & $1.43_{-0.15}^{+0.14}$ \\
J118.1364-26.0263 & 3.159168 & 0.076 & 0.081 & 14.366 & 14.258 & 0.14 & $0.174 \pm 0.015$ & $0.50_{-0.20}^{+0.18}$ & $-0.44_{-0.20}^{+0.18}$ \\
J118.1717-28.2232 & 1.836496 & 0.155 & 0.082 & 14.966 & 14.748 & 0.32 & $0.254 \pm 0.018$ & $1.86_{-0.16}^{+0.15}$ & $0.70_{-0.16}^{+0.15}$ \\
J118.6780-30.9643 & 1.914314 & 0.143 & 0.101 & 13.927 & 13.745 & 0.15 & $0.371 \pm 0.017$ & $1.66_{-0.10}^{+0.10}$ & $0.75_{-0.10}^{+0.10}$ \\
J120.3173-30.6071 & 1.635983 & 0.156 & 0.086 & 14.574 & 14.397 & 0.22 & $0.266 \pm 0.016$ & $1.60_{-0.13}^{+0.13}$ & $0.54_{-0.13}^{+0.13}$ \\
J121.7666-30.2091 & 1.623933 & 0.266 & 0.149 & 15.683 & 15.388 & 0.49 & $0.263 \pm 0.023$ & $2.61_{-0.20}^{+0.19}$ & $1.42_{-0.20}^{+0.19}$ \\
J122.5464-32.9936 & 4.368425 & 0.371 & 0.137 & 15.381 & 14.987 & 0.76 & $0.189 \pm 0.021$ & $1.53_{-0.25}^{+0.22}$ & $-0.16_{-0.25}^{+0.22}$ \\
J183.0555+26.0000 & 1.552371 & -0.278 & 0.115 & 14.449 & 14.676 & -0.70 & $0.118 \pm 0.034$ & $-0.06_{-0.74}^{+0.55}$ & $0.38_{-0.74}^{+0.55}$ \\
J275.6627-12.0194 & 2.767335 & 0.160 & 0.171 & 12.949 & 12.679 & 0.35 & $0.690 \pm 0.018$ & $1.98_{-0.06}^{+0.06}$ & $0.79_{-0.06}^{+0.06}$ \\
J311.5037+47.0071 & 2.701917 & 0.053 & 0.125 & 12.856 & 12.753 & 0.07 & $0.470 \pm 0.010$ & $1.15_{-0.05}^{+0.05}$ & $0.46_{-0.05}^{+0.05}$ \\
J322.5803+48.1366 & 3.722594 & 0.017 & 0.126 & 13.554 & 13.488 & -0.00 & $0.407 \pm 0.014$ & $1.56_{-0.07}^{+0.07}$ & $0.78_{-0.07}^{+0.07}$ \\
J336.4884+56.9915 & 2.408172 & 0.278 & 0.137 & 15.242 & 14.788 & 0.66 & $0.216 \pm 0.021$ & $1.65_{-0.23}^{+0.20}$ & $0.04_{-0.23}^{+0.20}$ \\
J340.1084+58.6023 & 2.775725 & 0.293 & 0.105 & 14.350 & 13.903 & 0.73 & $0.491 \pm 0.015$ & $2.54_{-0.07}^{+0.06}$ & $0.87_{-0.07}^{+0.06}$ \\
\enddata
\tablenotetext{a}{These IDs encode the star's approximate RA and Dec in decimal degrees, and allow lookup in the ATLAS DR1 database (see \citet{ATLASvar})}
\tablenotetext{b}{The ATLAS $c$ (cyan) and $o$ (orange) photometric bands are defined in \citet{Tonry2018}}
\tablenotetext{c}{Peak-to-trough amplitude based on Fourier fitting of the ATLAS data. Values provided are the average of $c$ and $o$ band amplitudes, but typically they were very similar.}
\tablenotetext{d}{Parallaxes are from Gaia EDR3 \citep{EDR3}.}
\tablenotetext{e}{The $g$, $r$, $z$, and $K$ magnitudes are from \citet{refcat} and represent a homogeneous recalibration of magnitudes drawn from Gaia DR2, Pan-STARRS, 2MASS, and other sources. $M_V$ and $M_K$ are derived from these magnitudes and the Gaia parallaxes.}
\tablenotetext{f}{The absolute magnitudes quoted for these stars are 3 $\sigma$ upper limits, since their nominal parallaxes are negative.}
\tablenotetext{g}{Gaia DR3 does not provide a parallax for this star, but we tentatively list it as a probable $\alpha^2$ CVn based on its color.}
\end{deluxetable*}

\startlongtable
\begin{deluxetable*}{llcccccccc}
\tablewidth{0pt}
\tabletypesize{\small}
\tablecaption{ATLAS UCBH stars that probably are not $\alpha^2$ CVn stars\label{tab:nocat}}
\tablehead{\colhead{ATLAS ID\tablenotemark{a}} & \colhead{Period (d)} & \colhead{$c-o$\tablenotemark{b}} & \colhead{amplitude\tablenotemark{c}} & \colhead{g} & \colhead{r} & \colhead{g-z} & \colhead{parallax\tablenotemark{d} (mas)} & \colhead{$M_V$\tablenotemark{e}}& \colhead{$M_K$\tablenotemark{e}}}
\startdata
J057.9558+54.6451 & 2.222517 & 0.469 & 0.083 & 13.811 & 13.195 & 1.03 & $0.843 \pm 0.039$ & $3.08_{-0.10}^{+0.10}$ & $0.98_{-0.10}^{+0.10}$ \\
J059.8990+50.2781 & 4.764477 & 0.576 & 0.215 & 16.508 & 15.729 & 1.37 & $0.282 \pm 0.039$ & $3.30_{-0.32}^{+0.28}$ & $0.72_{-0.32}^{+0.28}$ \\
J067.0419+51.6124 & 6.716061 & 0.413 & 0.193 & 16.158 & 15.558 & 0.98 & $0.209 \pm 0.034$ & $2.41_{-0.38}^{+0.33}$ & $0.59_{-0.38}^{+0.33}$ \\
J074.6123+26.0721 & 2.610292 & 0.395 & 0.078 & 14.988 & 14.503 & 0.92 & $0.375 \pm 0.021$ & $2.57_{-0.12}^{+0.12}$ & $0.69_{-0.12}^{+0.12}$ \\
J075.5525+46.9691 & 2.664824 & 0.440 & 0.095 & 14.993 & 14.414 & 0.98 & $0.366 \pm 0.025$ & $2.47_{-0.16}^{+0.15}$ & $0.43_{-0.16}^{+0.15}$ \\
J082.6906-06.8709 & 1.137648 & 1.094 & 0.215 & 16.911 & 15.651 & 2.76 & $2.814 \pm 0.028$ & $8.43_{-0.02}^{+0.02}$ & $3.93_{-0.02}^{+0.02}$ \\
J083.1858+21.5801 & 2.081992 & 0.750 & 0.398 & 16.055 & 15.071 & 1.78 & $0.787 \pm 0.059$ & $4.96_{-0.17}^{+0.16}$ & $1.42_{-0.17}^{+0.16}$ \\
J089.0960+24.8012 & 1.363957 & 0.369 & 0.164 & 15.985 & 15.468 & 0.83 & $0.256 \pm 0.036$ & $2.72_{-0.33}^{+0.29}$ & $0.74_{-0.33}^{+0.29}$ \\
J095.3593+12.9723 & 4.515769 & 0.521 & 0.213 & 16.424 & 15.589 & 1.19 & $0.225 \pm 0.040$ & $2.70_{-0.43}^{+0.36}$ & $0.47_{-0.43}^{+0.36}$ \\
J101.5867-01.4697\tablenotemark{g} & 1.903659 & 0.214 & 0.081 & 13.631 & 13.349 & 0.49 & $1.146 \pm 0.628$ & $3.76_{-1.72}^{+0.95}$ & $2.29_{-1.72}^{+0.95}$ \\
J114.4350-24.3432 & 2.242821 & 0.376 & 0.123 & 15.442 & 14.839 & 0.92 & $0.345 \pm 0.023$ & $2.78_{-0.15}^{+0.14}$ & $0.85_{-0.15}^{+0.14}$ \\
J138.5489+06.3771 & 4.097173 & 0.440 & 0.178 & 15.372 & 14.779 & 1.05 & $0.598 \pm 0.031$ & $3.91_{-0.12}^{+0.11}$ & $1.47_{-0.12}^{+0.11}$ \\
J207.7199+36.7006\tablenotemark{f} & 3.31534 & -0.373 & 0.318 & 13.290 & 13.645 & -0.95 & $1.095 \pm 0.024$ & $3.69_{-0.05}^{+0.05}$ & $4.00_{-0.05}^{+0.05}$ \\
J238.9224-20.7209 & 1.028827 & 1.235 & 0.260 & 15.358 & 13.964 & 3.11 & $7.149 \pm 0.019$ & $8.82_{-0.01}^{+0.01}$ & $3.72_{-0.01}^{+0.01}$ \\
J266.7656+06.0408\tablenotemark{g} & 4.555574 & 0.508 & 0.298 & 14.688 & 14.303 & 0.56 & $0.954 \pm 0.029$ & $4.36_{-0.07}^{+0.06}$ & \nodata \\
J279.0944-07.2749 & 3.236192 & 0.561 & 0.176 & 15.053 & 14.377 & 1.29 & $0.520 \pm 0.022$ & $3.24_{-0.09}^{+0.09}$ & $0.85_{-0.09}^{+0.09}$ \\
J280.0484-07.0582 & 2.216235 & 0.384 & 0.118 & 14.433 & 13.968 & 0.89 & $0.354 \pm 0.020$ & $1.91_{-0.13}^{+0.12}$ & $-0.11_{-0.13}^{+0.12}$ \\
J299.4473+32.3039 & 1.72322 & 0.477 & 0.132 & 15.495 & 14.875 & 1.08 & $0.241 \pm 0.020$ & $2.05_{-0.19}^{+0.18}$ & $-0.04_{-0.19}^{+0.18}$ \\
J304.4798+40.9564 & 3.372826 & 0.931 & 0.166 & 16.663 & 15.470 & 2.30 & $0.615 \pm 0.023$ & $4.91_{-0.08}^{+0.08}$ & $0.99_{-0.08}^{+0.08}$ \\
J306.8515+38.9758 & 1.381301 & 0.653 & 0.156 & 15.629 & 14.726 & 1.55 & $0.556 \pm 0.019$ & $3.83_{-0.07}^{+0.07}$ & $1.07_{-0.07}^{+0.07}$ \\
J313.8366+15.3471 & 3.093374 & 0.655 & 0.345 & 16.114 & 15.109 & 1.69 & $0.615 \pm 0.029$ & $4.47_{-0.11}^{+0.10}$ & $1.36_{-0.11}^{+0.10}$ \\
J315.9822+55.3700 & 2.644264 & 0.507 & 0.129 & 14.768 & 14.136 & 1.12 & $0.628 \pm 0.101$ & $3.39_{-0.38}^{+0.32}$ & $1.16_{-0.38}^{+0.32}$ \\
J319.0262+51.3139 & 2.200647 & 0.431 & 0.133 & 14.329 & 13.792 & 0.97 & $0.512 \pm 0.011$ & $2.56_{-0.05}^{+0.05}$ & $0.61_{-0.05}^{+0.05}$ \\
J331.2729+57.0606 & 4.63986 & 0.392 & 0.116 & 14.578 & 14.072 & 0.89 & $0.318 \pm 0.018$ & $1.80_{-0.12}^{+0.12}$ & $-0.17_{-0.12}^{+0.12}$ \\
J336.8577+57.3726 & 7.399448 & 0.757 & 0.160 & 14.127 & 13.122 & 1.79 & $4.654 \pm 0.472$ & $6.88_{-0.23}^{+0.21}$ & $3.40_{-0.23}^{+0.21}$ \\
\enddata
\tablenotetext{a}{These IDs encode the star's approximate RA and Dec in decimal degrees, and allow lookup in the ATLAS DR1 database (see \citet{ATLASvar})}
\tablenotetext{b}{The ATLAS $c$ (cyan) and $o$ (orange) photometric bands are defined in \citet{Tonry2018}}
\tablenotetext{c}{Peak-to-trough amplitude based on Fourier fitting of the ATLAS data. Values provided are the average of $c$ and $o$ band amplitudes, but typically they were very similar.}
\tablenotetext{d}{Parallaxes are from Gaia DR3 \citep{DR3}.}
\tablenotetext{e}{The $g$, $r$, $z$, and $K$ magnitudes are from \citet{refcat} and represent a homogeneous recalibration of magnitudes drawn from Gaia DR2, Pan-STARRS, 2MASS, and other sources. $M_V$ and $M_K$ are derived from these magnitudes and the Gaia parallaxes.}
\tablenotetext{f}{While most of the objects in this table are deemed too red to be Ap stars, this one is probably too blue.}
\tablenotetext{g}{This star is in the right color range to be a reddened Ap star, but it appears to be insufficiently luminous.}
\end{deluxetable*}

\begin{figure*} 
\plottwo{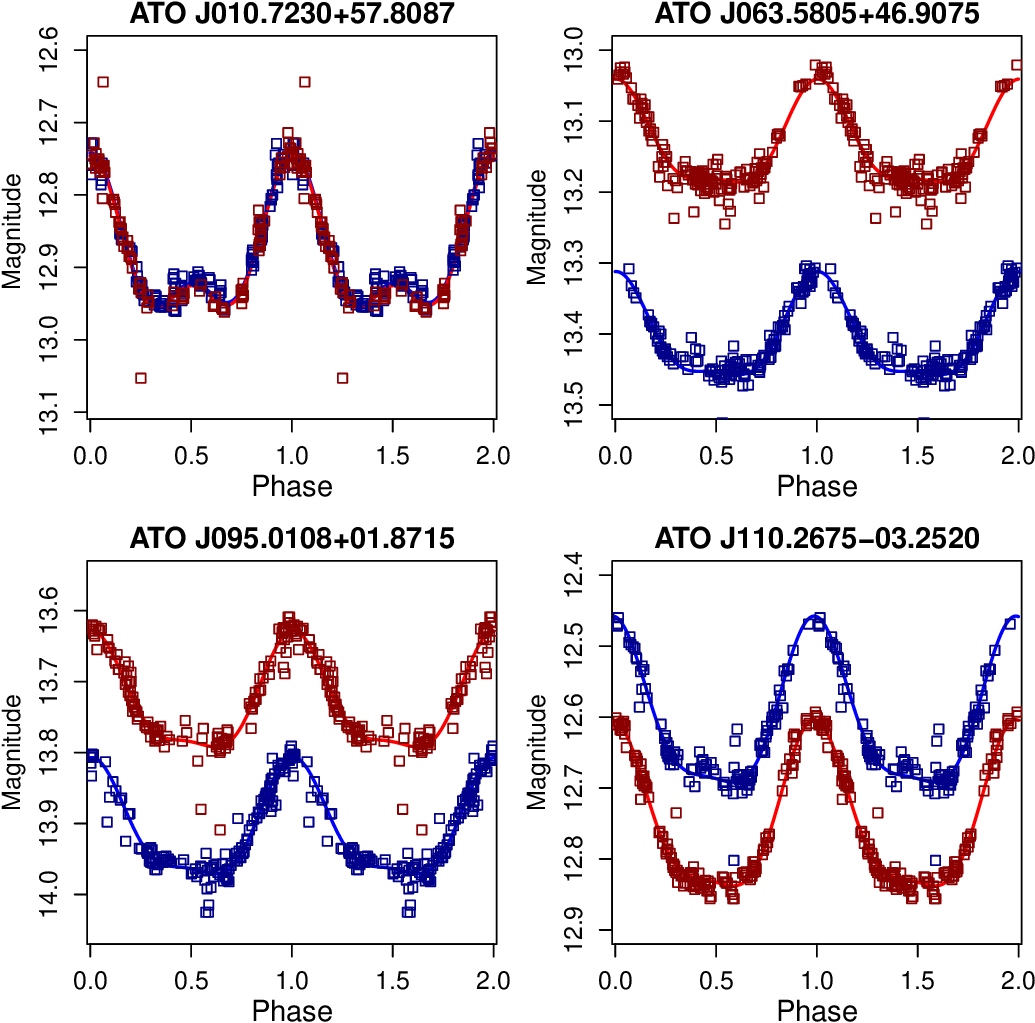}{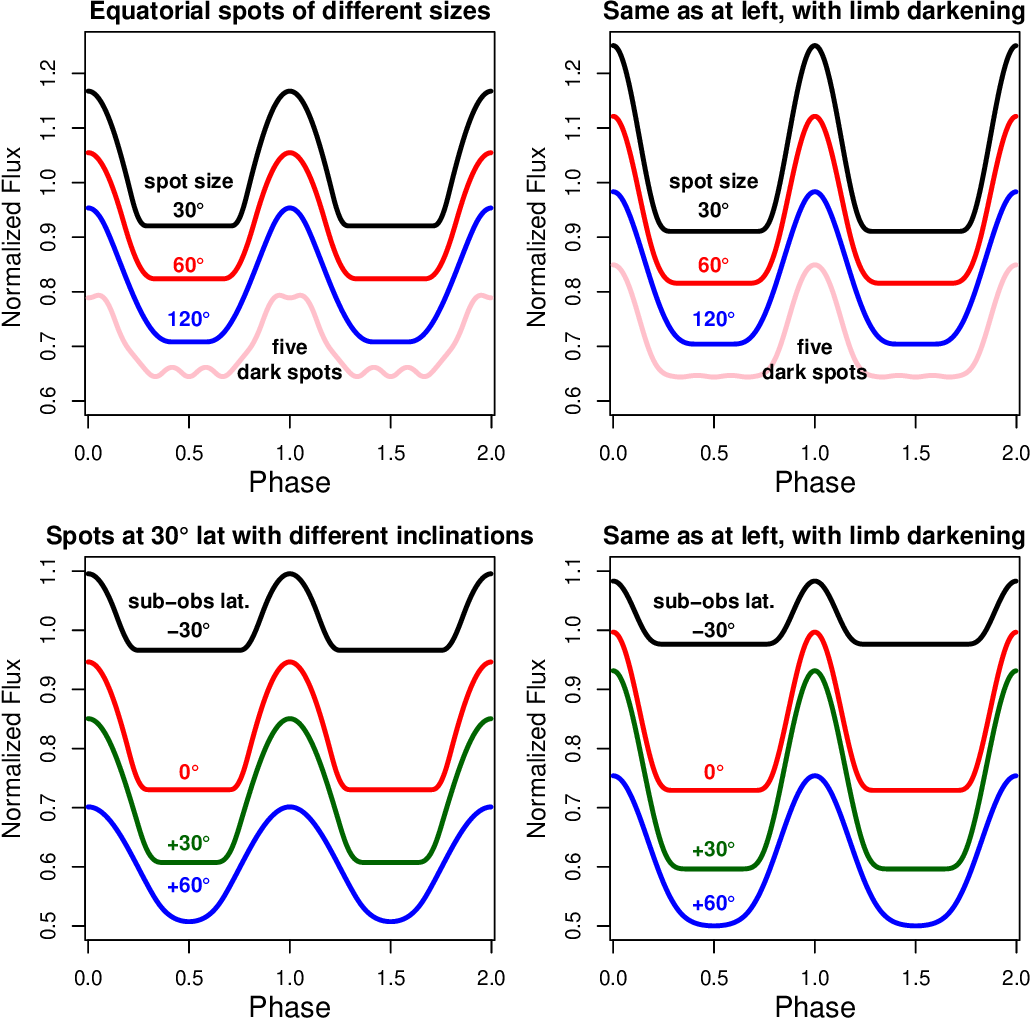}
\caption{The characteristic lightcurves of ATLAS UCBH stars match those expected from a single bright spot on a rotating star for a variety of spot sizes, contrasts, latitudes, and sub-observer latitudes, both with and without limb-darkening. {\em Left:} Example lightcurves for four ATLAS UCBH stars, with c-band photmetry in blue, o-band in red, and Fourier fits (see \citet{ATLASvar}) plotted as solid curves. {\em Right:} Model light curves for rotating stars with a single bright spot. They resemble UCBH light curves except for spot diameters larger than 120$^{\circ}$, and for high sub-observer latitudes (that is, low inclinations), when the spots can become circumpolar. A similar light curve results from the more contrived case of a ring of dark spots with a gap (pink curves in the upper plots).}
\label{fig:fourexamp} 
\end{figure*}

Figures \ref{fig:many} and \ref{fig:fourexamp} give examples of the specific and unusual `upside-down contact binary' light curve shape that defines these stars. The UCBH lightcurves have narrow, symmetrical maxima and broad, nearly-flat minima. The machine learning we used in ATLAS DR1 classified most UCBH stars as pulsators. However, known classes of pulsating variables typically have markedly asymmetrical maxima (the familiar `sawtooth' shape of RRAB and $\delta$ Scuti variables) or else more sinusoidal variations (RRC and some types of Cepheids). Furthermore, our spectral types combined with absolute magnitudes based on Gaia distances (Section \ref{sec:HR}) indicate that most UCBH variables are A-type or late B-type main sequence stars, and when such stars pulsate (e.g., the $\delta$ Scuti stars), their fundamental frequencies lead to periods much shorter than those of UCBH stars.

Alternatively, as illustrated by Figure \ref{fig:fourexamp}, a rotating star with a single bright spot near the equator will naturally exhibit UCBH-type variations for a wide range of non-polar viewing geometries. The probability that our line of sight to a randomly oriented star will be inclined by an angle $\theta$ to its rotation axis is proportional to $\sin(\theta)$, so near-polar ($\theta \sim 0$) viewing geometries are statistically disfavored. Hence, if the Milky Way contains a population of stars that have a single bright spot at low latitude, simple geometry dictates that the majority of them {\em must} appear as UCBH stars.

A bright, low-latitude spot is not the only way to produce a UCBH-type rotational lightcurve. The pink curves in Figure \ref{fig:fourexamp} demonstrate that a band of equatorial dark spots with a gap in it will produce a similar effect. However, this band-and-gap explanation (though it may apply to some systems) is more complex and specific: Occam's Razor favors the model with just a single bright spot. Unless stars with such a feature are vanishingly rare in the Milky Way, the geometrical argument we have already made demonstrates that they must be represented among our UCBH objects.

\begin{figure*} 
\plottwo{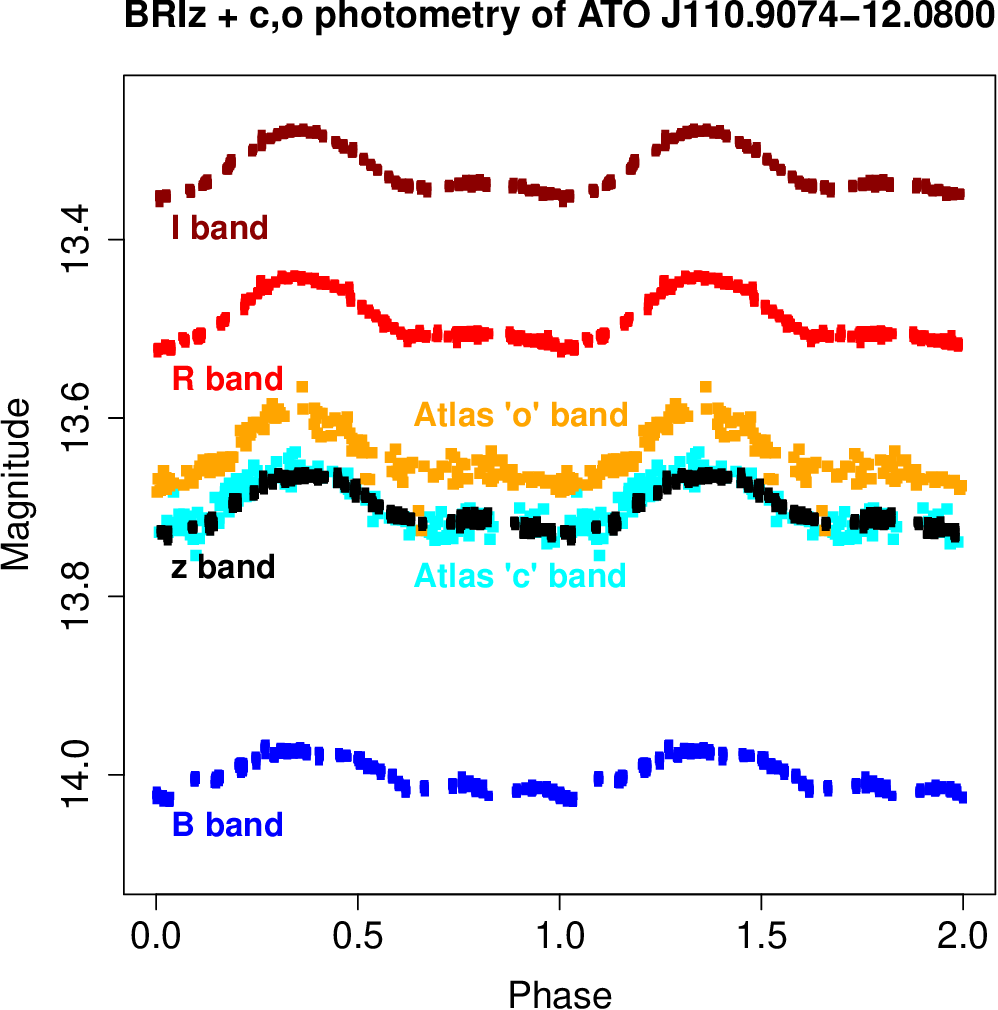}{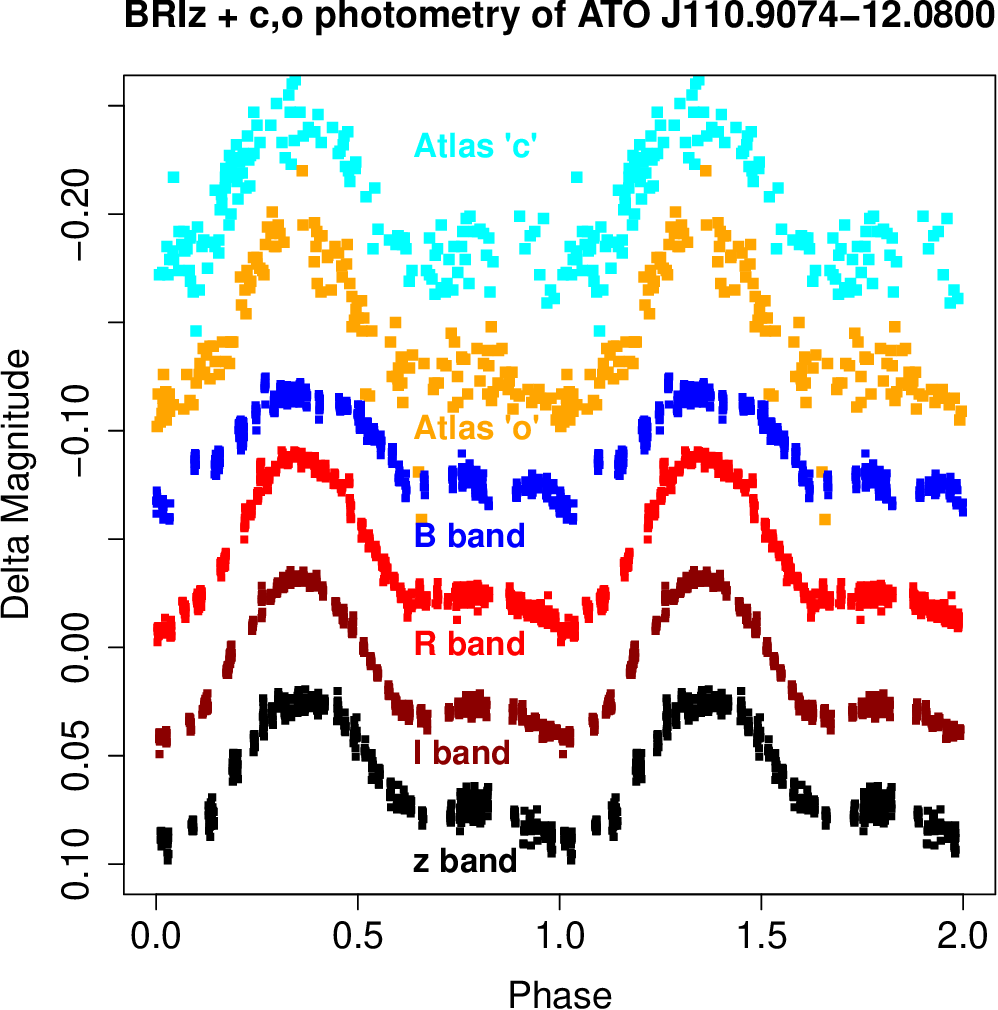}
\caption{Folded lightcurves of ATO J110.9074-12.0800. {\em Left:} apparent magnitude vs. phase for targeted $B$, $R$, $I$, $z$-band photometry from the University of Hawaii 2.2m telescope on Mauna Kea, together with the $o$ and $c$-band photometry from ATLAS. {\em Right:} Same data as at left, but with magnitude offsets applied to facilitate comparing the light curves in greater detail. The light curve shape is consistent from 2015 (ATLAS data) through 2019 (UH 2.2m data), and across the different photometric bands probed here --- in strong contrast to the sinusoidally-varying $\alpha^2$ CVn variable KIC 2569073, which showed a phase-reversal in the $B$-band relative to $R_{\mathrm{C}}$ and $I_{\mathrm{C}}$ \citep{Drury2017}.}
\label{fig:UH88}
\end{figure*}

To explore the photometric behavior of UCBH stars with higher precision and more wavelength bands, we monitored the UCBH star ATO J110.9074-12.0800 intensively for five nights (UT 2019 January 18--22) using the University of Hawaii 2.2 meter telescope on Maunakea (Figure \ref{fig:UH88}). This star was later spectrally confirmed to be an Ap star and hence an $\alpha^2$ CVn variable. For our photometric monitoring we used the $B$, $R$, $I$, and $z$ filters, finding very similar lightcurve shape in all filters, with slightly reduced amplitudes in the $B$ and possibly $z$ bands.  Interestingly, we do not see a phase-reversal at $B$-band relative to $R$ and $I$, such as was noted by \citet{Drury2017} in the sinusoidally-varying $\alpha^2$ CVn variable KIC 2569073.

The phase and lightcurve shape of ATO J110.9074-12.0800 have remained coherent and unchanging to within measurement error from the beginning of ATLAS data aquisition in October 2015 up through the UH 2.2 meter monitoring in January 2019. In the higher-precision 2.2 meter data the maximum continues to appear very symmetrical. A slant and slight `bump' on the floor of the broad minimum, hinted at in the ATLAS data, are confirmed by the more precise photometry. Such features also seem to be indicated in ATLAS data for other UCBH stars, notably ATO J010.7230+57.8087 (Figure \ref{fig:fourexamp}). This indicates that the feature producing the photometric maximum, while dominant, is not necessarily the only photospheric inhomogeneity on a typical UCBH star. 

\section{Spectra of UCBH Stars} \label{sec:spec}

\subsection{Observations} \label{sec:specobs}

In 2018 we acquired spectra of five ATLAS UCBH stars with the GMOS spectrograph on the 8-meter Gemini North telescope under proposal ID GN-2018B-Q-216. This proposal was designed to take advantage of the worst usable weather by targeting bright objects that could be usefully observed even through moonlit cloud with bad seeing. This observing plan produced a win-win situation in which we provided Gemini queue observers with targets for conditions when almost nothing else could be observed, while the spectra they acquired for us were in fact considerably better than our nominal requirements. This happened because worst-case observing conditions are statistically rare, so the majority of our data were acquired in somewhat better weather than we had planned for (though still too poor for most observing programs).

Our plan of exploiting the worst usable weather at Gemini determined determined both our choice of target objects (the brightest ATLAS UCBH stars observable) and the slit width (2.0 arcsec, to allow for very bad seeing). We used the GMOS B1200 grating, which delivers a nominal resolution of $R=3744$ at 4630\AA~ with an 0.5 arcsecond slit\footnote{\url{https://www.gemini.edu/instrumentation/gmos/components\#Gratings}}. Our 2.0 arcsecond slit would therefore be expected to deliver $R=936$, four times worse than nominal -- but the actual resolution could be higher if the seeing was smaller than the slit. We observed each target alternately with two different central wavelength settings, 4400\AA~ and 4680\AA, enabling us to fill in gaps between the three GMOS CCDs and obtain continuous spectral coverage from 3650--5500\AA.

In addition to our GMOS observations, we acquired spectra of nine additional UCBH stars using the SNIFS instrument \citep{SNIFS} on the University of Hawaii 2.2 meter telescope on Maunakea, in February and March 2019. A substantially larger number of spectra was originally expected from this observing program, but it was plagued with bad weather and equipment problems. The SNIFS instrument has a blue module delivering spectral coverage from 3200--5600\AA~ with resolution $R\sim1000$ at 4300\AA, and a red module covering 5200-10000\AA~ with $R\sim1300$ at 7600\AA~ \citep{SNIFS}.

Our nominal resolution element at $\sim4500$\AA~ should be $4500/936$ = 4.8\AA~ with GMOS and $4500/1000$ = 4.5\AA~ with SNIFS. Comparisons of our GMOS and SNIFS spectra demonstrate that GMOS has actually delivered better resolution --- an indication that the seeing was smaller than our 2.0 arcsecond slit width during our GMOS observations.

\subsection{Identification of Ap Stars} \label{sec:specan}

\begin{figure*} 
\plottwo{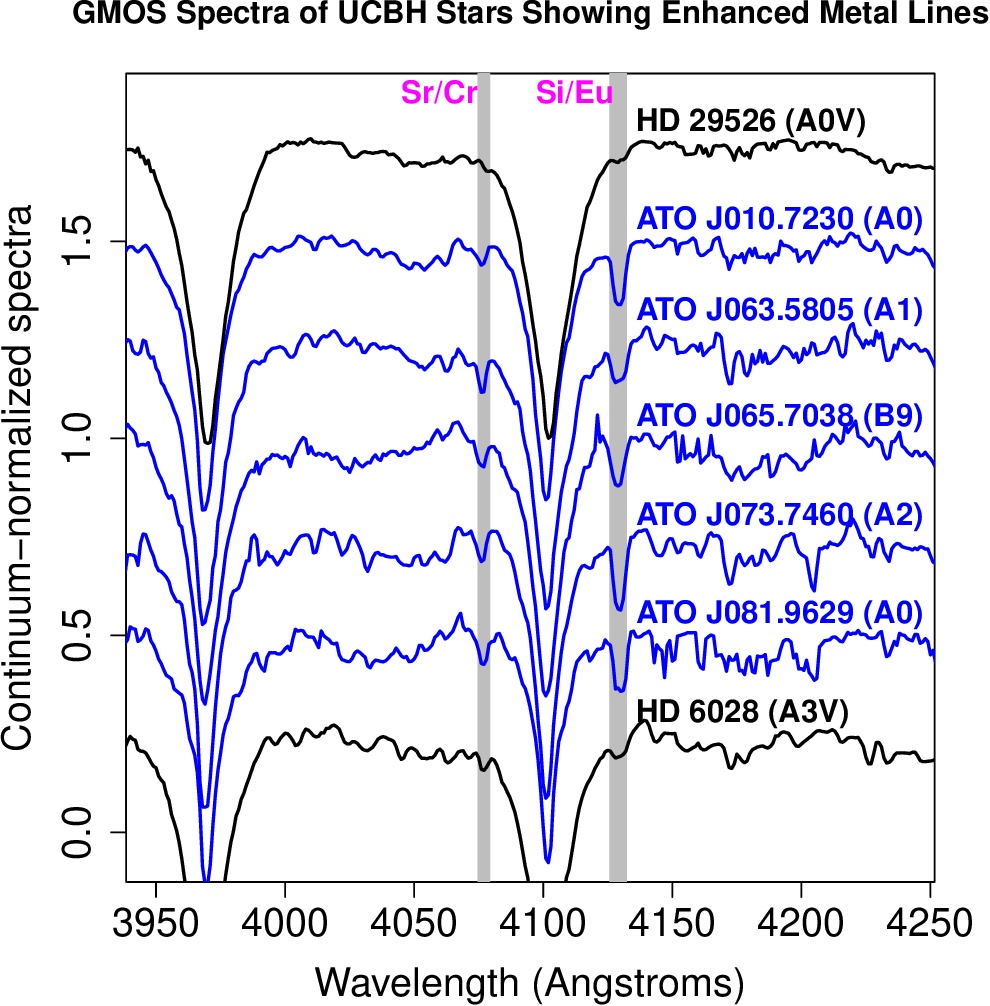}{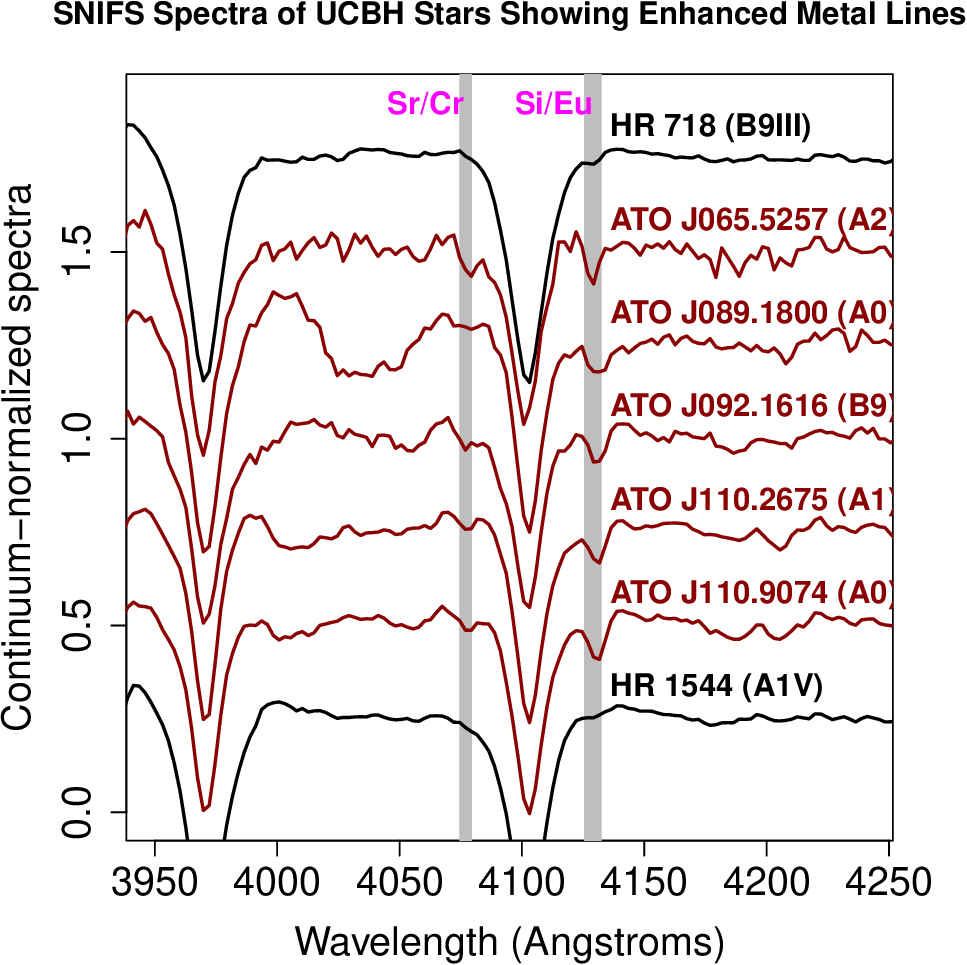}
\caption{{\em Left:} Low-resolution spectra of five ATLAS `UCBH' stars acquired with Gemini/GMOS (blue), compared with those of normal A-type standard stars (black). The UCBH stars have strong enhancements of specific heavy elements in their atmospheres, as indicated by the lines labeled Sr/Cr and Si/Eu. As the lines are blended at this resolution, the relative contributions of the different elements cannot be determined. {\em Right:} Similar comparison for spectra of five additional UCBH stars (dark red) acquired with the SNIFS spectrograph at the UH 2.2 meter telescope on Mauna Kea. Although the SNIFS spectra do not have as high resolution and SNR as those from GMOS, the peculiar metal lines can still be clearly seen.}
\label{fig:UCBH} 
\end{figure*}

Our Gemini spectra (Figure \ref{fig:UCBH}) showed all five science targets to be Ap/Bp stars: that is, A-type or B-type stars with enormously enhanced abundances of a few specific heavy elements (mainly silicon, europium, chromium, and strontium). The peculiar lines that we detect most strongly form two blends, one near 4080\AA~ (likely a blend of Sr and Cr) and another near 4130\AA~ (a blend of Si and Eu). The resolution of our spectra is insufficient to determine the relative contributions of each element to the blended lines. Previous work on such stars \citep[see, e.g.][]{Preston1974,Dukes2018} distinguishes fine gradations of spectral classification depending on magnetic field strength and on what elements are enhanced to what extent. Since multiple lines are blended in our spectra, they do not enable us to assign exact types of chemical peculiarity --- but they do establish that our targets fall into the broad category of chemically peculiar A-type or B-type stars.

The SNIFS spectra, though not matching the resolution and signal-to-noise ratio of GMOS, are nevertheless sufficient to show that five of our SNIFS targets are Ap/Bp stars, while the rest are not A or B stars at all. Figure \ref{fig:UCBH} shows spectra for all ten of our spectrally confirmed Ap/Bp stars, five from GMOS and five from SNIFS.

\subsection{Spectral Types of UCBH stars} \label{sec:spectype}

We have attempted to determine spectral types for the UCBH stars for which we have spectra. We have done this classification manually using comparison spectra from the Stellar Spectral Flux Library of \citet{Pickles1998}, guided in part by the diagnostic spectral lines mentioned in the Atlas of Stellar Spectra\footnote{\url{https://ned.ipac.caltech.edu/level5/ASS_Atlas/frames.html}}. The effective resolution of our SNIFS spectra matches fairly well with the 5\AA~sampling used in the \citet{Pickles1998} library. The higher resolution of our GMOS spectra made narrow spectral lines look too deep relative to the \citet{Pickles1998} library, so we smoothed the GMOS classification spectra using a Gaussian blur of $\sigma=3$\AA.

\begin{figure*} 
\plottwo{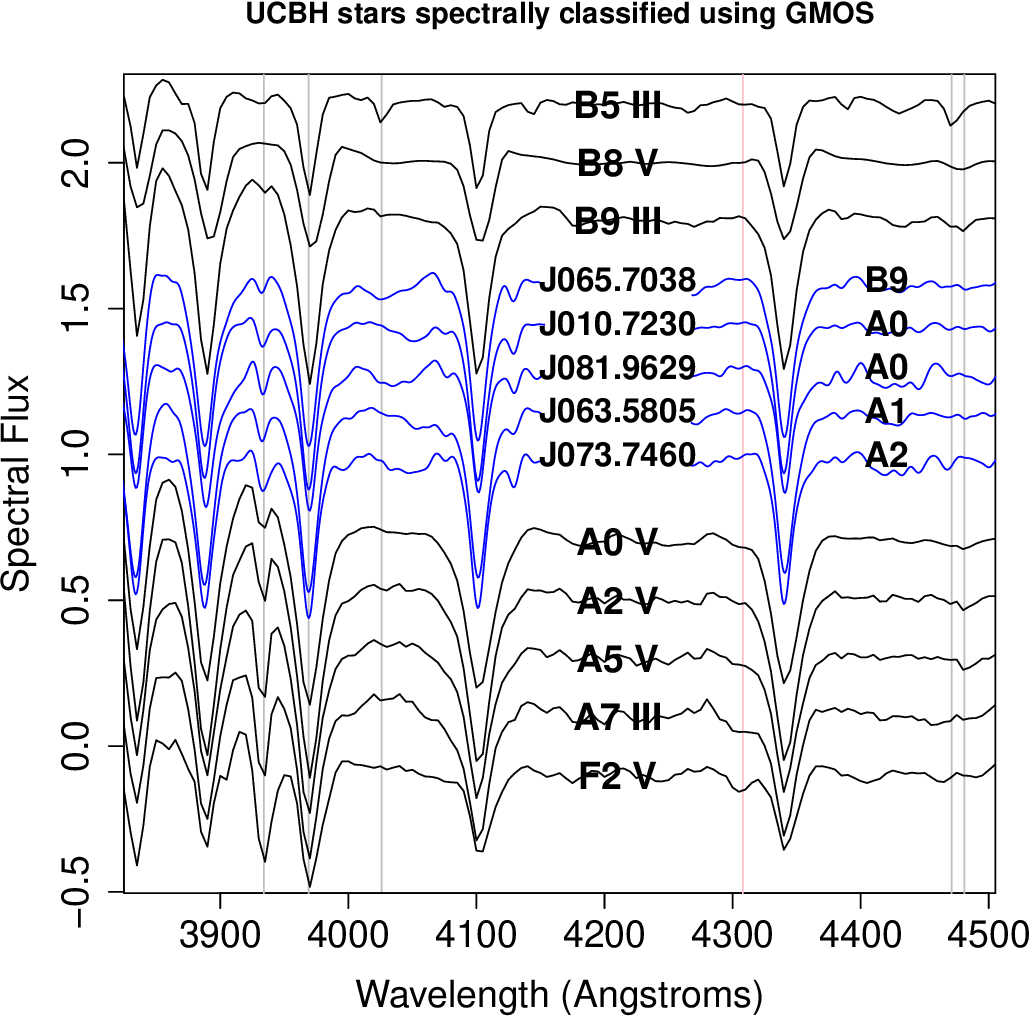}{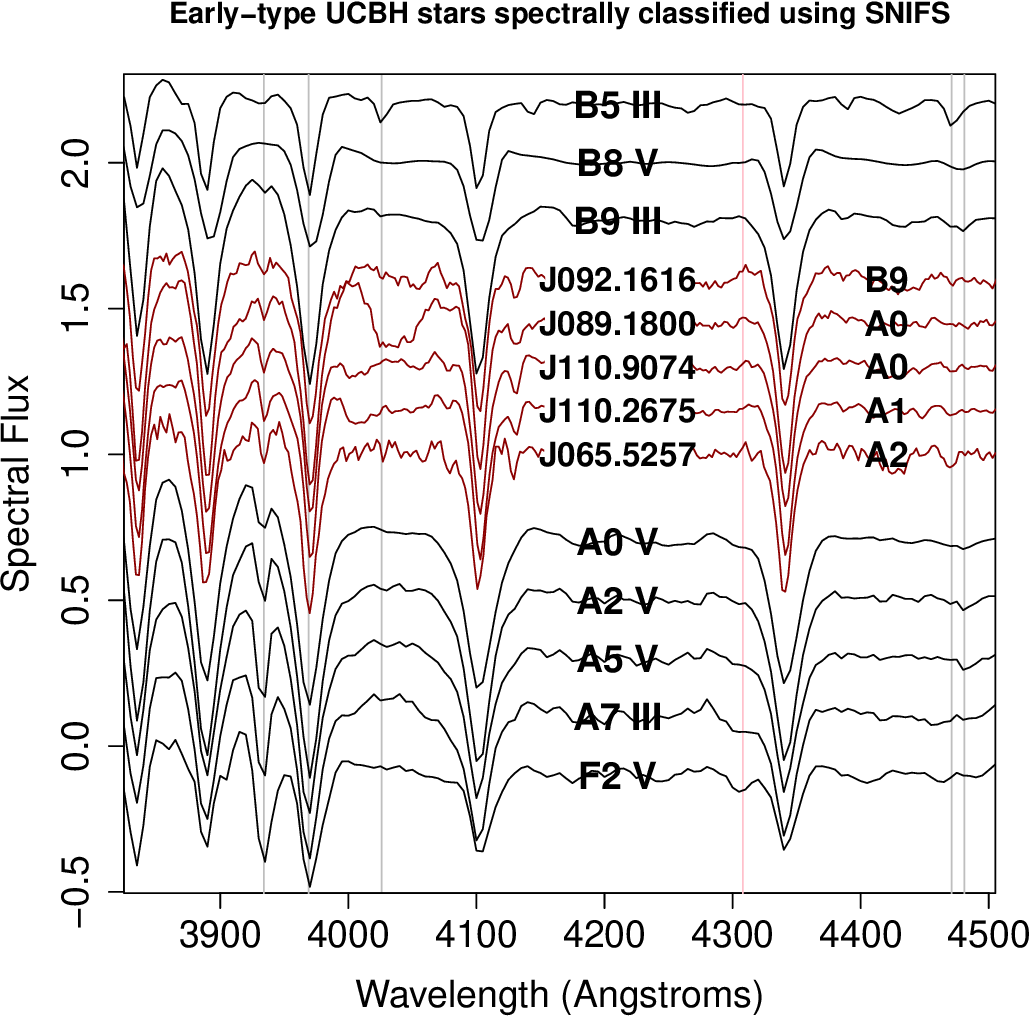}
\caption{Spectral classifications of A-type UCBH stars with GMOS (left, blue) and SNIFS (right, dark red). Comparison spectra, plotted in black, are from the library of \citet{Pickles1998}. Our GMOS spectra have been smoothed to match the library resolution.  Gray vertical lines mark some of the spectral lines mentioned as useful for classification in the Atlas of Stellar Spectra (see text). The spectral types we found for these stars, based almost exclusively on the changing strength of the calcium H line at 3969\AA, are written on the spectra in these plots and listed Table \ref{tab:spectypes}.}
\label{fig:specclass} 
\end{figure*}

The classification spectra of our $\alpha^2$ CVn stars, with appropriate comparison spectra from \citet{Pickles1998}, are shown in Figure \ref{fig:specclass}, and the spectral types we assigned are in Table \ref{tab:spectypes}. For these A-type (or very late B) stars, classification with uncertainty no greater than one spectral subtype appears to be possible based on the strength of the calcium H line at 3969\AA~ (which is the only one in our spectra with significant diagnostic power). Based on this, we would expect our spectral types to be quite accurate --- with the important caveat that the chemical peculiarity of our stars might have affected the calcium H line or our perception of it (e.g., by changing the nearby continuum). That the spectra are not typical of A stars is obvious even at reduced resolution: besides numerous lines not present in the comparison spectra, the hydrogen Balmer lines seem somewhat weaker in the UCBH stars. However, unless classification biases from the peculiar spectra are extremely severe, there is no doubt that all of our $\alpha^2$ CVn stars are early A or very late B-type. 

\begin{figure*} 
\plottwo{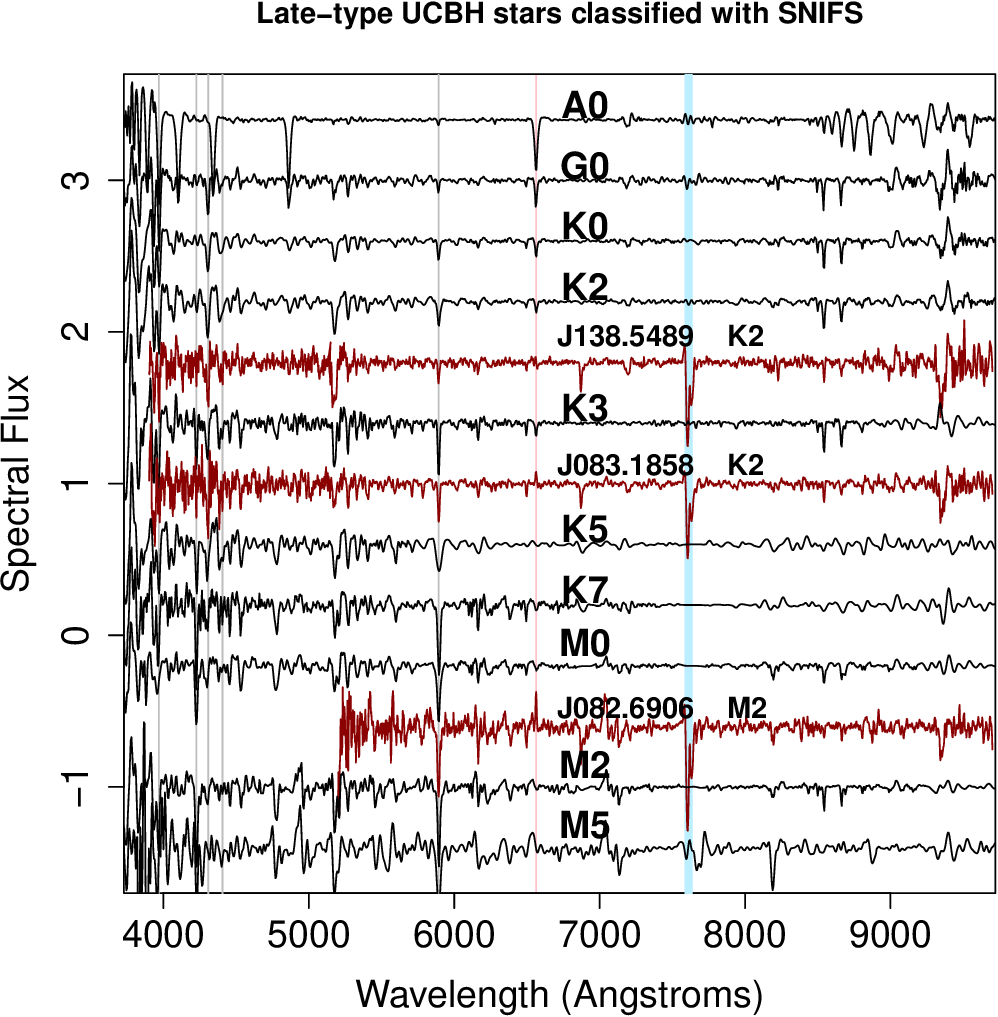}{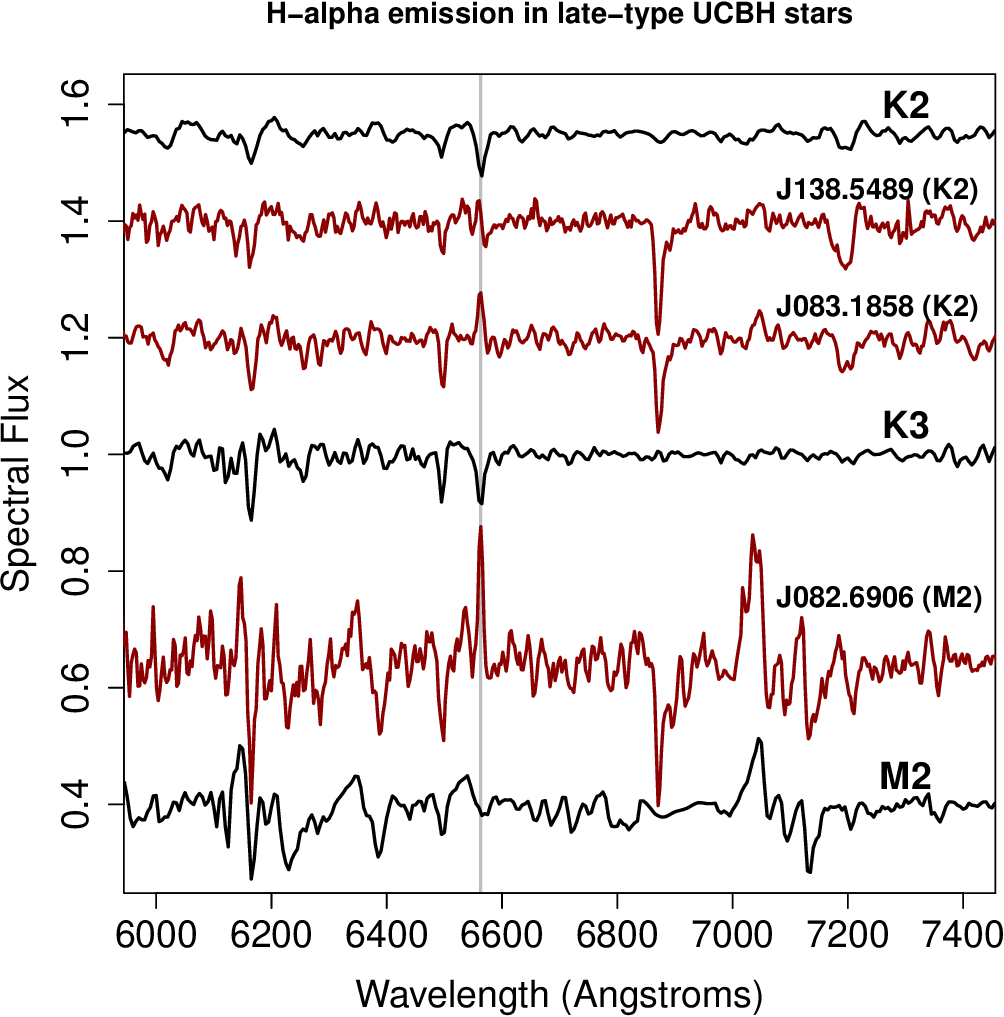}
\caption{Spectral classifications of late-type UCBH stars with SNIFS (left), and the detection of H$\alpha$ emission in these stars (right, with H$\alpha$ marked by a gray vertical line). The target spectra are shown in dark red, while comparison spectra from the library of \citet{Pickles1998} are plotted in black. Gray vertical lines mark some of the spectral lines mentioned as useful for classification in the Atlas of Stellar Spectra, but as the Atlas covers only relatively short wavelengths, we have used many other lines and bands to arrive at the spectral types given in Table \ref{tab:spectypes}. The pale blue line near 7600\AA~in the left-hand plot marks the Fraunhofer A band, which is not intrinsic to the stars but is caused by oxygen in Earth's atmosphere.}
\label{fig:specclass2} 
\end{figure*}

Four of the UCBH stars for which we obtained SNIFS spectra were not $\alpha^2$ CVn stars. One of these, ATO J207.7199+36.7006, is in fact the known subdwarf OB star PG 1348+369 \citep{Green1986,Wesemael1992}. Since our spectra are consistent with the published results, and indicate a star much too hot to be an $\alpha^2$ CVn variable, we have not attempted to reclassify this object. Spectra of the remaining three UCBH stars, which have much later spectral types, are shown along with \citet{Pickles1998} comparison spectra in Figure \ref{fig:specclass2}. The spectral types we assigned them are provided in Table \ref{tab:spectypes}. For these classifications, the diagnostic lines listed in the Atlas of Stellar Spectra were of limited value because our SNIFS spectra of these red objects were very faint in the blue region covered by the Atlas. Hence, we made use of many other lines at much longer wavelengths that appeared to be diagnostic based on their variations with spectral type seen in the \citet{Pickles1998} library. We expect our classifications of these later-type stars to have an accuracy of around two spectral subtypes. Interestingly, all three of our late-type UCBH stars show significant H$\alpha$ emission (Figure \ref{fig:specclass2}, right panel).

\begin{deluxetable}{lccc}
\tablewidth{0pt}
\tabletypesize{\small}
\tablecaption{ATLAS UCBH stars classified with low-resolution spectra\label{tab:spectypes}}
\tablehead{ & \colhead{Spectral} & & \\ \colhead{Star} & \colhead{Type} & \colhead{Instrument} & \colhead{$\alpha^2$ CVn?}}
\startdata
ATO J010.7230+57.8087 & A0 & GMOS & yes \\
ATO J063.5805+46.9075 & A1 & GMOS & yes \\
ATO J065.5257+51.2992 & A2 & SNIFS & yes \\
ATO J065.7038+47.6938 & B9 & GMOS & yes \\
ATO J073.7460+43.3008 & A2 & GMOS & yes \\
ATO J081.9629+42.4325 & A0 & GMOS & yes \\
ATO J082.6906-06.8709 & M2 & SNIFS & no \\
ATO J083.1858+21.5801 & K2 & SNIFS & no \\
ATO J089.1800+11.3598 & A0 & SNIFS & yes \\
ATO J092.1616+30.8849 & B9 & SNIFS & yes \\
ATO J110.2675-03.2520 & A1 & SNIFS & yes \\
ATO J110.9074-12.0800 & A0 & SNIFS & yes \\
ATO J138.5489+06.3771 & K2 & SNIFS & no \\
ATO J207.7199+36.7006 & sd0 & SNIFS & no \\
\enddata
\end{deluxetable}

\section{UCBH stars from Bernhard et al. (2020)} \label{sec:bernhard}

\citet{Bernhard2020} make a remarkable contribution to the photometry of $\alpha^2$ CVn stars by analyzing photometry of 294 bright Ap/Bp stars with previous spectroscopic identifications. They use data from three surveys that, by using very small apertures, maintain photometric precision for bright stars that are saturated in ATLAS photometry. Examining their published lightcurves, we identify 33 UCBH stars, which we list in Table \ref{tab:bernhard} together with relevant parameters for these stars from \citet{Bernhard2020}, \citet{refcat}, and Gaia DR3 \citep{DR3}. These stars, being much brighter than the ATLAS UCBH stars of Table \ref{tab:yescat}, can more easily be explored with high resolution, high-SNR spectroscopy or spectropolarimetry such as is required for detailed abundance analysis or Zeeman Doppler imaging. Many of the \citet{Bernhard2020} stars might be less interesting targets because they have much smaller photometric amplitudes relative to the ATLAS UCBH stars --- but a few exceptions (particularly HD 191287, HD 77314, and HD 205938) have perfect UCBH lightcurves with ATLAS-like amplitudes. As these stars are magnitudes brighter than any in the ATLAS catalog, they are the most promising targets for followup spectroscopy and Zeeman Dopper imaging to probe the chemical abundances and magnetic field topologies of $\alpha^2$ CVn stars with UCBH lightcurves (see Section \ref{sec:conc}).

\startlongtable
\begin{deluxetable*}{llccccccccl}
\tablewidth{0pt}
\tabletypesize{\small}
\tablecaption{$\alpha^2$ CVn stars with UCBH lightcurves from \citet{Bernhard2020} \label{tab:bernhard}}
\tablehead{\colhead{Star} & \colhead{Period(d)} & \colhead{Sp\tablenotemark{a}} & \colhead{amplitude\tablenotemark{b}} & \colhead{V\tablenotemark{c}} & \colhead{g-z\tablenotemark{d}} & \colhead{parallax\tablenotemark{e} (mas)} & \colhead{$M_V$\tablenotemark{f}}& \colhead{$M_K$\tablenotemark{f}} & Remarks\tablenotemark{g}}
\startdata
HD 7546 & 3.9725 & A0 & 0.03 & 9.43 & -0.525 & $2.9380 \pm 0.1027$ & $1.77_{-0.08}^{+0.08}$ & $-1.00_{-0.08}^{+0.08}$ &   \\
HD 26792 & 3.8023 & B8 & 0.04 & 6.69 & -0.581 & $6.1745 \pm 0.0307$ & $0.64_{-0.01}^{+0.01}$ & $0.54_{-0.01}^{+0.01}$ & Strong \\
HD 30466 & 1.40687 & A0 & 0.03 & 7.25 & -0.402 & $5.1396 \pm 0.2791$ & $0.80_{-0.12}^{+0.11}$ & $0.37_{-0.12}^{+0.11}$ &   \\
HD 39317 & 2.6558 & B9 & 0.01 & 5.59 & -0.731 & $6.8553 \pm 0.0959$ & $-0.23_{-0.03}^{+0.03}$ & $-0.25_{-0.03}^{+0.03}$ &   \\
HD 43819 & 14.981 & B9 & 0.02 & 6.27 & -0.796 & $4.0642 \pm 0.1735$ & $-0.69_{-0.09}^{+0.09}$ & $-0.54_{-0.09}^{+0.09}$ &   \\
HD 44903 & 1.41143 & A5 & 0.03 & 8.37 & -0.445 & $4.7409 \pm 0.0430$ & $1.75_{-0.02}^{+0.02}$ & $1.46_{-0.02}^{+0.02}$ & Strong \\
HD 46462 & 10.346 & B9 & 0.06 & 7.53 & -0.935 & $4.1051 \pm 0.3830$ & $0.60_{-0.21}^{+0.19}$ & $0.87_{-0.21}^{+0.19}$ &   \\
HD 51418 & 5.4377 & A0 & 0.13 & 6.67 & -0.536 & $5.6092 \pm 0.0929$ & $0.42_{-0.04}^{+0.04}$ & $0.35_{-0.04}^{+0.04}$ & Strong \\
HD 55667 & 1.79690 & A2 & 0.03 & 6.95 & -0.645 & $7.4800 \pm 0.0277$ & $1.32_{-0.01}^{+0.01}$ & $1.31_{-0.01}^{+0.01}$ & Strong \\
HD 56273 & 1.78678 & B8 & 0.04 & 7.90 & -0.749 & $2.7056 \pm 0.0353$ & $0.06_{-0.03}^{+0.03}$ & $0.22_{-0.03}^{+0.03}$ &   \\
HD 77314 & 2.86445 & A2 & 0.08 & 7.24 & -0.528 & $4.4294 \pm 0.0493$ & $0.47_{-0.03}^{+0.02}$ & $0.20_{-0.03}^{+0.02}$ & Ideal \\
HD 88701 & 25.77 & B9 & 0.06 & 9.30 & -0.453 & $2.0931 \pm 0.0201$ & $0.90_{-0.02}^{+0.02}$ & $0.82_{-0.02}^{+0.02}$ &   \\
HD 129189 & 1.35563  & B9 & 0.03 & 8.61 & -0.511 & $3.6861 \pm 0.0231$ & $1.44_{-0.01}^{+0.01}$ & $1.22_{-0.01}^{+0.01}$ &   \\
HD 142884 & 0.80296 & B9 & 0.02 & 6.77 & -0.581 & $5.7423 \pm 0.0415$ & $0.56_{-0.01}^{+0.02}$ & $0.47_{-0.01}^{+0.02}$ &   \\
HD 150714 & 1.62906 & A0 & 0.05 & 7.56 & 0.342 & $6.0507 \pm 0.0331$ & $1.47_{-0.01}^{+0.01}$ & $0.96_{-0.01}^{+0.01}$ &   \\
HD 151199 & 2.2267 & A3 & 0.01 & 6.17 & -0.539 & $9.6547 \pm 0.1270$ & $1.09_{-0.03}^{+0.03}$ & \nodata &   \\
HD 154187 & 8.096 & A0 & 0.03 & 9.27 & 0.518 & $3.3825 \pm 0.0349$ & $1.92_{-0.02}^{+0.02}$ & $0.33_{-0.02}^{+0.02}$ &   \\
HD 173650 & 9.976 & A0 & 0.04 & 6.51 & -0.456 & $4.0447 \pm 0.0261$ & $-0.46_{-0.01}^{+0.01}$ & $-0.57_{-0.01}^{+0.01}$ &   \\
HD 176582 & 1.58193 & B5 & 0.02 & 6.40 & -0.643 & $3.2506 \pm 0.0411$ & $-1.04_{-0.03}^{+0.03}$ & $-0.52_{-0.03}^{+0.03}$ &   \\
HD 177410 & 1.12318 & B9 & 0.03 & 6.50 & -0.959 & $5.1153 \pm 0.0366$ & $0.04_{-0.01}^{+0.02}$ & $0.42_{-0.01}^{+0.02}$ &   \\
HD 184020 & 2.5515 & A0 & 0.02 & 8.16 & -0.448 & $5.0175 \pm 0.0359$ & $1.66_{-0.01}^{+0.02}$ & $1.62_{-0.01}^{+0.02}$ &   \\
HD 184905 & 1.84548 & A0 & 0.04 & 6.61 & -0.709 & $5.0323 \pm 0.0301$ & $0.12_{-0.01}^{+0.01}$ & $0.23_{-0.01}^{+0.01}$ &   \\
HD 191287 & 1.62342 & B9 & 0.18 & 8.17 & -0.569 & $3.4259 \pm 0.0301$ & $0.84_{-0.02}^{+0.02}$ & $0.93_{-0.02}^{+0.02}$ & Ideal \\
HD 195447 & 5.3970 & B9 & 0.03 & 7.57 & -0.492 & $2.2576 \pm 0.0563$ & $-0.66_{-0.06}^{+0.05}$ & $-0.77_{-0.06}^{+0.05}$ &   \\
HD 196542 & 1.7929 & A4 & 0.02 & 9.04 & -0.550 & $2.6255 \pm 0.0162$ & $1.14_{-0.01}^{+0.01}$ & $0.66_{-0.01}^{+0.01}$ &   \\
HD 205938 & 8.335 & B9 & 0.05 & 6.46 & -0.617 & $4.4302 \pm 0.0326$ & $-0.31_{-0.02}^{+0.02}$ & $-0.11_{-0.02}^{+0.02}$ & Ideal \\
HD 207188 & 2.6735 & A0 & 0.06 & 7.66 & -0.858 & $3.4180 \pm 0.0467$ & $0.33_{-0.03}^{+0.03}$ & $0.61_{-0.03}^{+0.03}$ &   \\
HD 213871 & 1.95070 & B9 & 0.05 & 7.38 & -0.536 & $3.5779 \pm 0.0663$ & $0.15_{-0.04}^{+0.04}$ & $0.13_{-0.04}^{+0.04}$ & Strong \\
HD 220668 & 6.1606 & A0 & 0.09 & 7.64 & -0.564 & $2.4295 \pm 0.0277$ & $-0.43_{-0.03}^{+0.02}$ & $-0.38_{-0.03}^{+0.02}$ &   \\
HD 221394 & 2.8600 & A0 & 0.04 & 6.39 & -0.456 & $7.0059 \pm 0.0333$ & $0.62_{-0.01}^{+0.01}$ & $0.54_{-0.01}^{+0.01}$ & Strong \\
HD 223660 & 2.8258 & B8 & 0.03 & 8.09 & -0.507 & $2.2318 \pm 0.0396$ & $-0.17_{-0.04}^{+0.04}$ & $-0.12_{-0.04}^{+0.04}$ &   \\
HD 224166 & 3.5139 & B9 & 0.02 & 6.93 & -0.565 & $2.9238 \pm 0.0404$ & $-0.74_{-0.03}^{+0.03}$ & $-0.62_{-0.03}^{+0.03}$ &   \\
TYC 2850-263-1 & 12.440 & A & 0.01 & 9.79 & -0.038 & $3.9682 \pm 0.0189$ & $2.78_{-0.01}^{+0.01}$ & $1.60_{-0.01}^{+0.01}$ &   \\
\enddata
\tablenotetext{a}{Periods and spectral types are from \citet{Bernhard2020}}
\tablenotetext{b}{$V$-band peak-to-trough variability amplitude (magnitudes) from \citet{Bernhard2020}}
\tablenotetext{c}{Average $V$-band from \citet{Bernhard2020}}
\tablenotetext{d}{From \citet{refcat}}
\tablenotetext{e}{Parallaxes are from Gaia DR3 \citep{DR3}.}
\tablenotetext{f}{$M_V$ and $M_K$ are derived from \citet{refcat} photometry and the Gaia parallaxes.}
\tablenotetext{g}{Suitability for high resolution spectroscopic investigation, based on visual examination of the \citet{Bernhard2020} light curve: `Strong' means a good candidate with amplitude much larger than the photometric scatter. `Ideal' means, additionally, high-amplitude variations that perfectly exemplify the UCBH lightcurve shape.}
\end{deluxetable*}

\section{HR Diagrams of UCBH Stars} \label{sec:HR}

\begin{figure*} 
\plottwo{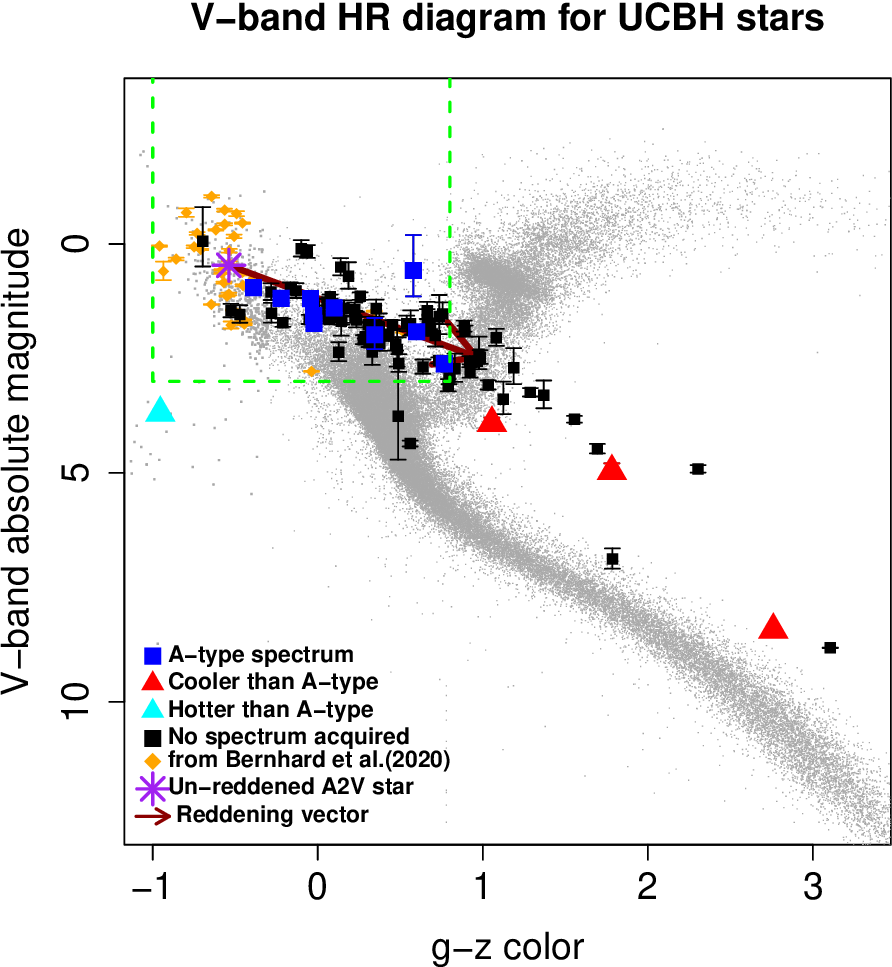}{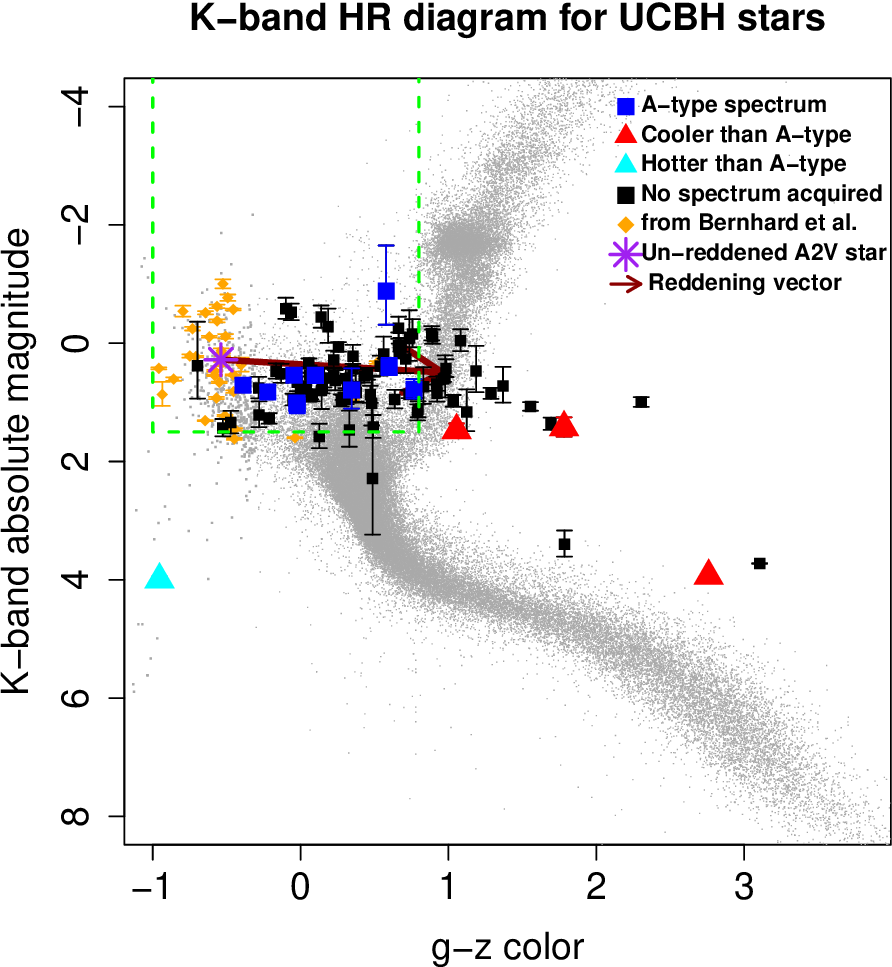}
\caption{HR diagrams for UCBH stars for V-band absolute magnitude (left) and K-band (right). Absolute magnitudes of UCBH stars are based on fluxes from \citet{refcat} and parallaxes from Gaia DR3 \citep{Gaia,EDR3,DR3}. Green rectangles illustrate the regions of each diagram from which objects were selected as probable $\alpha^2$ CVn stars for inclusion in Table \ref{tab:yescat}. The dark red arrow in each figure indicates the direction a star moves as it becomes increasingly dust-reddened. While classification is not definitive without spectra, the vast majority ($>90$\%) of stars within the green rectangles, as well as some objects that lie outside of them but along the reddening vectors, should be $\alpha^2$ CVn objects. The distribution of UCBH stars from \citet{Bernhard2020} is consistent with the expectation that these nearer objects should be less reddened. Small gray points illustrate the Galactic field population using data from \citet{refcat}.  }
\label{fig:CMD} 
\end{figure*}

The precision and comprehensive sky coverage of Gaia parallaxes \citep{Gaia} are revolutionizing Galactic stellar astrophysics, and our UCBH stars are no exception. Figure \ref{fig:CMD} shows observers' HR diagrams of our UCBH stars against a background plot of about $10^5$ high Galactic latitude stars which outline the main sequence and the giant branch. We used $g-z$ colors to obtain strong wavelength leverage and reduce sensitivity to the known photometric variability of these stars. Magnitudes are taken from \citet{refcat}, where we have determined $V$ magnitudes from $g$ and $r$ using Equation \ref{eq:Vmag}, which comes from a transformation derived by Robert Lupton\footnote{\url{http://classic.sdss.org/dr4/algorithms/sdssUBVRITransform.html}}. This transformation should be valid through the whole range of stellar colors and spectral types relevant to this paper, since it is based on Peter Setson's photometric standard stars\footnote{See, e.g., \url{https://www.cadc-ccda.hia-iha.nrc-cnrc.gc.ca/en/community/STETSON/}}, which span $B-V$ colors ranging from -0.4 to +3.5 mag: i.e., the entire range of ordinary stars from spectral types O and B through late-M.

\begin{equation} \label{eq:Vmag}
V = g - 0.5784*(g - r) - 0.0038
\end{equation}

For the UCBH stars, we have used parallaxes from Gaia Data Release 3 \citep[DR3;][]{DR3}, while for the gray background points in Figure \ref{fig:CMD} we have used the parallaxes in the photometric catalog of \citet{refcat}, which come from Gaia Data Release 2 \citep[DR2;][]{DR2}.

Figure \ref{fig:CMD} shows a large range of colors even for the UCBH stars that we have confidently determined to be A or late B-type $\alpha^2$ CVn stars. Furthermore, they are mostly redder and less luminous than nearby main sequence stars with early A spectral types. To determine if this can be plausibly attributed to dust reddening and extinction, we used the interstellar extinction coefficients provided in Table 21.6 of \citet{aaq} for $R_V=3.1$. Since this table does not provide coefficients for the $g$ and $z$ bands we chose for our colors, we interpolated it to the effective wavelengths given for these bands by \citet{Bessel2005}. Hence, we arrived at interstellar extinction coefficients (relative to the $V$ band) of 1.2426 for $g$, 0.4930 for $z$, and 0.108 for $K$. From these, we calculated the reddening vectors plotted in both panels of Figure \ref{fig:CMD}. These vectors indicate the direction a star moves on the figure as it becomes increasingly dust-reddened. We set the origin of each vector at the position of an un-reddened A2V star, intended to be characteristic of a `typical' $\alpha^2$ CVn star unaffected by dust extinction --- hence, we expect reddened stars of A or late B-type to fall along the reddening vector in each plot.

\begin{figure} 
\includegraphics[width=3.5in]{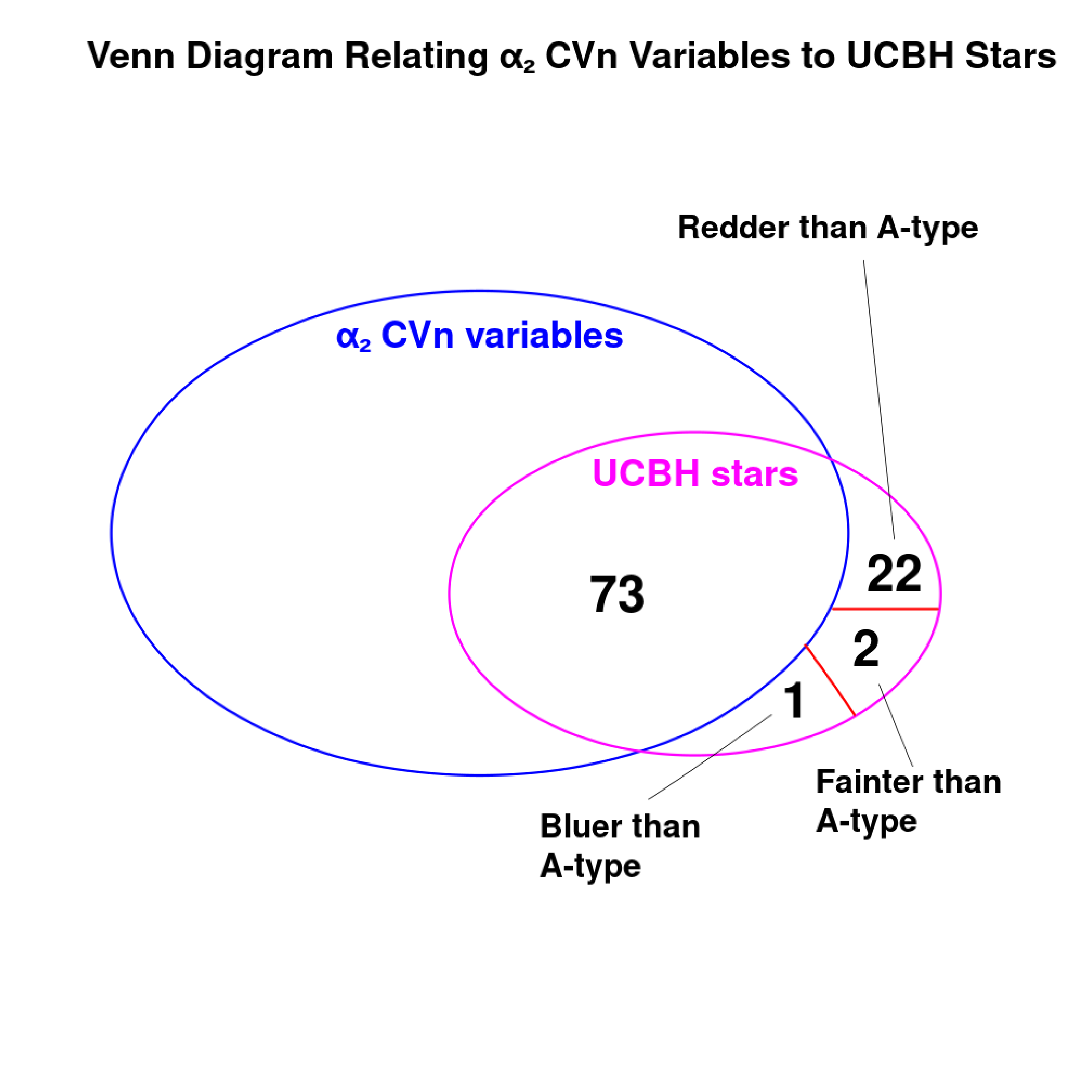}
\caption{Venn diagram illustrating that most UCBH stars appear to be $\alpha^2$ CVn variables, although only a minority of known $\alpha^2$ CVn variables have UCBH-type lightcurves. Since we have spectra for only fourteen out of 98 UCBH stars, the dividing line between $\alpha^2$ CVn variables and either hotter or cooler UCBH stars is based on simple color cuts on the $o-c$ color obtained from ATLAS lightcurves. Hence, the counts are very approximate and could be affected by interstellar reddening, both here and in Tables \ref{tab:yescat} and \ref{tab:nocat} where there same color cuts have been used.}
\label{fig:Venn} 
\end{figure}

The reddening vectors plotted in Figure \ref{fig:CMD} indicate that our $\alpha^2$ CVn UCBH stars all have colors and absolute magnitudes close to what would be expected for reddened main sequence stars of early A-type (or late B-type). The amount of interstellar extinction implied varies greatly from star to star, but approaches two magnitudes at $V$ band for our reddest spectrally confirmed $\alpha^2$ CVn stars. By contrast, the much nearer sample of UCBH stars from \citet{Bernhard2020} are consistent with A or late B-type stars with near-zero dust extinction --- as we should expect given their much smaller distances relative to the ATLAS UCBH stars. There may be an indication that the ATLAS UCBH stars are slightly underluminous (they tend to lie slightly below the line of the reddening vector), but we cannot conclude this with confidence given our rather simplistic reddening correction.

We have used Figure \ref{fig:CMD} as a guide to the range of color and absolute magnitude inhabited by UCBH stars that are also A or late-B type $\alpha^2$ CVn variables. We have selected the range -1.0 to 0.8 in $g-z$ color, and absolute magnitude thresholds of 3.0 for $M_V$ and 1.5 for $M_K$, indicated by green dashed rectangles in Figure \ref{fig:CMD}. We believe most of the UCBH stars in these regions of the HR diagrams will also be $\alpha^2$ CVn variables. Interlopers are possible, for example from less-reddened and slightly overluminous objects of later spectral types. However, the fact that no such interlopers were identified among our spectral sample of ten objects suggests they will be a small minority. Similarly, Figure \ref{fig:CMD} indicates that many of the UCBH stars redder than our limit of $g-z = 0.8$ would also be perfectly consistent with strongly reddened $\alpha^2$ CVn variables. Some of them almost certainly are exactly that. However, objects beyond our red limit could also be less reddened evolved stars ascending the giant branch, and hence we believe $g-z = 0.8$ is a good provisional limit to maintain a fairly pure sample in the absence of spectra for most of the stars. The green rectangles drawn in Figure \ref{fig:CMD} therefore mark the boundary between stars listed as probable $\alpha^2$ CVn variables in Table \ref{tab:yescat} and those listed as probably something else in Table \ref{tab:nocat}. Figure \ref{fig:Venn} gives a Venn Diagram of the respective classifications.

\section{Discussion and Conclusion} \label{sec:conc}

Using photometry from the ATLAS survey \citep{Tonry2018}, we have identified a rare population of periodic variable stars (the `UCBH' stars) with characteristic lightcurves having broad minima, narrow, symmetrical maxima, and periods mostly in the range of 1--10 days. Among 142 million distinct stars analyzed in ATLAS DR1 \citep{ATLASvar}, only 98 are identified as UCBH stars. Though the relatively low amplitudes of the UCBH stars mean we could have identified them only among the brightest $\sim20$\% of the DR1 sample, the fact that we found fewer than 100 in all shows they are extremely rare. Our spectroscopy of these objects indicates that most ($\sim75$\%) of them are $\alpha^2$ CVn variables --- that is, Ap/Bp stars that show rotationally modulated photometric variations. Although most UCBH stars are $\alpha^2$ CVn variables, only a minority (10--15\%) of $\alpha^2$ CVn variables appear to be UCBH stars. Meanwhile, $\alpha^2$ CVn stars themselves are only a subset of the Ap/Bp stars, which in turn comprise a small fraction of all A-type and B-type stars.

We have demonstrated that a single bright feature at low latitude on a rotating star will produce a UCBH-type lightcurve, by geometrical necessity (Figure \ref{fig:fourexamp}), for the most probable viewing inclinations. Hence, if the Milky Way contains a non-negligible population of stars with bright low-latitude features, they should be represented among our UCBH objects. The fact that most UCBH stars are $\alpha^2$ CVn variables suggests that the dominant astrophysical effect that can produce such a feature is connected to the $\alpha^2$ CVn stars --- i.e., to the peculiar abundances of heavy elements that characterize them. Before discussing the physical connection between $\alpha^2$ CVn variables and UCBH stars in more detail in Section \ref{sec:yes}, we briefly consider the UCBH stars that do {\em not} fall into the $\alpha^2$ CVn class.

\subsection{UCBH stars that are NOT $\alpha^2$ CVn variables} \label{sec:no}

The single localized bright spot that most simply explains a UCBH-type lightcurve could be produced by phenomona not related to the $\alpha^2$ CVn stars. One example is an accretion stream impacting a stellar photosphere. This may be the explanation for some UCBH stars --- notably the hot subdwarf PG 1348+369 \citep{Green1986,Wesemael1992}.

An approximately even longitudinal distribution of {\em dark} star spots with a prominent gap could also produce a UCBH-type rotationally modulated lightcurve (Figure \ref{fig:fourexamp}), with the `missing' dark spots of the gap functioning like a single bright feature. While $\alpha^2$ CVn stars are too hot for the ordinary form of dark magnetic starspots, this scenario may apply to the significant minority (22 out of 98 in the ATLAS sample; see Figure \ref{fig:Venn}) of UCBH stars with much later spectral types --- a hypothesis which is further bolstered by the detection of strong H$\alpha$ emission in all three of the late-type UCBH stars for which we have spectra. Such emission is characteristic of late type stars that are magnetically active and heavily spotted.

While late-type UCBH stars exist, the variability we observe in A-type or B-type UCBH stars cannot reasonably be attributed to an unresolved late-type companion. Such a companion would have to be several times fainter than the primary to escape detection in our spectra, implying a very large-amplitude photometric variation for the late-type star itself. If late-type stars commonly had high-amplitude UCBH-type lightcurves, they should be much easier to detect in the field than as binary companions to brighter A stars that would dilute their photometric amplitudes. In this case, isolated late-type stars, being far more numerous than A-type or B-type stars, should dominate our UCBH sample --- the opposite of what we observe. Additionally, it would be a strange coincidence if late-type UCBH companions were found only around chemically peculiar A-type primaries. Furthermore, the literature contains many examples of UCBH-type variations in $\alpha^2$ CVn stars (Section \ref{sec:ACVNintro}), and these stars are known to exhibit correlated spectral variations that clearly implicate the Ap/Bp star itself --- rather than a hypothetical late-type companion --- as the photometric variable.

In short, although a tiny minority of late-type stars do exhibit UCBH-type lightcurves, there is no doubt that the UCBH variations we observe in Ap/Bp stars originate from the bright stars themselves and not from an unresolved late-type companion.

\subsection{UCBH stars that ARE $\alpha^2$ CVn variables} \label{sec:yes}

A large majority (73 out of 98) of the ATLAS UCBH stars appear to be $\alpha^2$ CVn variables --- that is, Ap/Bp stars with rotationally modulated photometric variability. The Ap/Bp stars are chemically peculiar A-type or B-type stars with greatly enhanced abundances of specific heavy elements in their photospheres. The enhanced abundances are believed to be caused by radiative levitation of the heavy elements in question \citep{Michaud1970}, which is strongly influenced (and likely enabled) by magnetic fields \citep{Michaud1981}.

The rotationally modulated variability of $\alpha^2$ CVn variables results from inhomogeneous distributions of the radiatively levitated elements across the stars' photospheres \citep{Shulyak2010}. For such a star, the single bright feature implied by a UCBH lightcurve is naturally interpreted as the region where the concentration of radiatively levitated heavy elements is the highest. Such a region is bright at optical wavelengths because the strong UV absorption lines of the levitated elements redistribute the star's intense UV flux into the optical \citep{Michaud1981}. This single region of greatly enhanced heavy element abundance likely owes its existence to a particular configuration of the magnetic field.

The expected causal connection between the magnetic field and the rotational light curve implies that the $\alpha^2$ CVn variables that share UCBH-type light curves may also have similar magnetic field topologies. In this context, it is interesting that the lightcurves of some $\alpha^2$ CVn UCBH stars show a small `bump' or secondary maximum in the center of the broad, nearly flat minimum. These include ATO J010.7230+57.8087, ATO J110.9074-12.0800, and others in the ATLAS sample --- and also examples from the literature such as HD 207188 \citep{Hensberge1977} and HD 54118 \citep{Catalano1993}. A photometric `bump' of this type would naturally be produced by a secondary bright spot at the antipode of the one responsible for the primary maximum. The greatly reduced photometric signature of this antipodal spot could be an effect of latitude. For example, if the two bright features were at stellar latitudes of $+30^{\circ}$ and $-30^{\circ}$, respectively, and the sub-observer latitude were $+30^{\circ}$, we would observe the primary bright feature to pass directly across the center of the stellar disk, producing a maximal photometric signature, while the antipodal feature would barely come into view and would create a greatly reduced photometric `bump' half a rotation later --- exactly what we observe in some objects. However, the fact that many lightcurves do {\em not} show a secondary photometric `bump' suggests that latitude is not the only effect: if antipodal spots are always present, they must be much smaller/fainter than the primary spot in many cases.

Particularly since there is sometimes evidence of an antipodal spot, it is tempting to conclude that the UCBH lightcurves imply a simple magnetic topology such as a dipole field. However both theory \citep{Michaud1981} and Zeeman Dopper observations \citep{Kochukhov2010,Kochukhov2015} point to a complex relationship between magnetic field configurations and inhomogeneity in the heavy element enhancements. For example, some elements concentrate where the magnetic field lines are vertical and some where they are horizontal \citep{Michaud1981}. Hence, conclusions that the UCBH stars have simple magnetic fields --- or even that they all have the same magnetic topology --- may be unwarranted.

Nevertheless, the shared UCBH-type light curve shape of many $\alpha^2$ CVn stars does suggest some commonality in the magnetic field distribution. This could make them interesting targets for Zeeman Doppler imaging and elemental abundace mapping --- especially since the ATLAS UCBH stars have photometric amplitudes larger than average for $\alpha^2$ CVn variables, likely indicating large and easy-to-measure spatial variations in the magnetic field and elemental abundances. Since the ATLAS UCBH stars are several magnitudes fainter than typical targets of Zeeman Doppler imaging, we have sought and identified 33 UCBH stars among the much brighter sample of $\alpha^2$ CVn variables published by \citet{Bernhard2020}. Most of these have substantially smaller photometric amplitudes than the ATLAS UCBH stars, but there are exceptions. The most promising of these --- bright, high-amplitude variables with perfect UCBH light curves --- are HD 191287, HD 77314, and HD 205938 \citep[see Table \ref{tab:bernhard} and ][]{Bernhard2020}. These stars are ideal targets for Zeeman Doppler imaging and other forms of high-resolution spectroscopic investigation to probe the detailed astrophysics behind the UCBH-type light curves of $\alpha^2$ CVn variables.

\section{Acknowledgments} 
This publication presents discoveries made by the Asteroid Terrestrial-Impact Last Alert System (ATLAS). Support for the ATLAS survey is provided by NASA grants NN12AR55G and 80NSSC18K0284 under the guidance of Lindley Johnson and Kelly Fast.

This research is based on observations obtained at the international Gemini Observatory, a program of NSF's NOIRLab, which is managed by the Association of Universities for Research in Astronomy (AURA) under a cooperative agreement with the National Science Foundation on behalf of the Gemini Observatory partnership: the National Science Foundation (United States), National Research Council (Canada), Agencia Nacional de Investigaci\'{o}n y Desarrollo (Chile), Ministerio de Ciencia, Tecnolog\'{i}a e Innovaci\'{o}n (Argentina), Minist\'{e}rio da Ci\^{e}ncia, Tecnologia, Inova\c{c}\~{o}es e Comunica\c{c}\~{o}es (Brazil), and Korea Astronomy and Space Science Institute (Republic of Korea). These observations were obtained under Gemini Program ID GN-2018B-Q-216.

This work was enabled by observations made from the Gemini North telescope and the University of Hawaii 2.2 meter telescope, both located within the Maunakea Science Reserve and adjacent to the summit of Maunakea. We are grateful for the privilege of observing the Universe from a place that is unique both for its astronomical quality and for its place in Hawaiian indigenous culture.

We thank Simon Murphy for helping us realize that our mysterious objects were  $\alpha^2$ CVn stars, and for giving us guidance about which elements were likely responsible for the peculiar spectral lines we observed.

This work has made use of data from the European Space Agency (ESA) mission
{\it Gaia} (\url{https://www.cosmos.esa.int/gaia}), processed by the {\it Gaia}
Data Processing and Analysis Consortium (DPAC,
\url{https://www.cosmos.esa.int/web/gaia/dpac/consortium}). Funding for the DPAC
has been provided by national institutions, in particular the institutions
participating in the {\it Gaia} Multilateral Agreement.

This publication makes use of the SIMBAD online database,
operated at CDS, Strasbourg, France, and the VizieR online database (see \citet{vizier}).

This publication makes use of data products from the Two Micron All Sky Survey, 
which is a joint project of the University of Massachusetts and the Infrared Processing
and Analysis Center/California Institute of Technology, funded by the National Aeronautics
and Space Administration and the National Science Foundation.

We have also made extensive use of information and code from \citet{nrc}.

Facilities: \facility{Gemini North, UH88}


\begin{thebibliography}{}
\bibitem[Bessell(2005)]{Bessel2005} Bessell, M. S. 2005, \araa, 43, 293
  %standard wavelengths for the SDSS filters.
\bibitem[Bernhard et al.(2020)]{Bernhard2020} Bernhard, K., H\"{u}mmerich, S., \& Paunzen, E. 2020, \mnras, 493, 3293
    %Amazing list of lightcurves, etc. from ASAS-3, KELT, and MASCARA data. 294 stars
\bibitem[Bernhard et al.(2021)]{Bernhard2021} Bernhard, K., H\"{u}mmerich, S., Paunzen, E. \& Sup\'{i}kov\'{a}, J. 2021, \mnras, 506, 4561
\bibitem[Catalano \& Leone (1993)]{Catalano1993} Catalano, F. A. \& Leone, F. 1993, \aaps, 97, 501
\bibitem[Chambers et al.(2016)]{PS1A} Chambers, K. C., Magnier, E. A., Metcalfe, N., et al. 2016, arXiv:1612.05560
%PS1 Surveys, Chambers, K.C., et al. 
\bibitem[Cox(2000)]{aaq} Cox, A. N. 2000, Allen's Astrophysical Quantities (Fourth Edition; New York, NY: Springer-Verlag New York, Inc.)
\bibitem[Drury et al.(2017)]{Drury2017} Drury, J. A., Murphy, S. J., Derekas, A., S\'{o}dor, \'{A}., Stello, D., Kuehn, C. A., Bedding, T. R., Bogn\'{a}r, Z., Szigeti, L., Szak\'{a}ts, R., S\'{a}rneczky, K., \& Moln\'{a}r, L. 2017, \mnras, 471, 3193
  %Kepler and ground-based photometry (0.6-m telescope in Hungary), and
  %Nordic Optical Telescope spectroscopy of Kepler Ap star KIC 2569073.
  %P=14.67 days, Peak-to-peak amplitudes are 0.13, 0.03, 0.07, 0.28, and 0.34
  %magnitudes in B, V, Rc, Ic, and Kp respectively. There is a phase-reversal
  %between B and V.
\bibitem[Dukes \& Adelman (2018)]{Dukes2018} Dukes, R. J. \& Adelman, S. J. 2018, \pasp, 130, 044202
%Photometry of eight ACVN stars. One (HD 26792) shows a UCBH-type lightcurve.
\bibitem[Flewelling et al.(2016)]{Flewelling2016} Flewelling, H., Magnier, E., Chambers, K. et al.  2016,
  arXiv:1612.05243
\bibitem[Gaia Collaboration et al.(2016)]{Gaia} Gaia Collaboration, Prusti, T., de Bruijne, J. H. J., et al. 2016, \aap, 595, A1
  %Overall Gaia paper
\bibitem[Gaia Collaboration et al.(2018)]{DR2} Gaia Collaboration, Brown, A. G. A., Vallenari, A., Prusti, T., de Bruijne, J. H. J., Babusiaux, C., \& Bailer-Jones, C. A. L. 2018, arXiv e-prints, pp. arXiv:1804.09365
\bibitem[Gaia Collaboration et al.(2021)]{EDR3} Gaia Collaboration, Brown, A. G. A., Vallenari. A., et al. 2021, \aap, 649, 1
%Gaia EDR3 reference. 
\bibitem[Gaia Collaboration et al.(2022)]{DR3} Gaia Collaboration, Vallenari, A., Brown, A. G. A., et al. 2022, arXiv e-prints, pp. arXiv:2208.00211
  %Main Gaia DR3 reference.
\bibitem[Graham et al.(2018)]{ZTF} Graham, M. et al. 2018, AAS meeting \#231, 354.16
%ZTF
\bibitem[Green et al.(1986)]{Green1986} Green, R. F., Schmidt, M., \& Liebert, J. 1986, \apj, 61, 305
%Reference for hot subdwarf ATO J207.7199+36.7006 = PG 1348+369.
\bibitem[Heinze et al.(2018)]{ATLASvar} Heinze, A. N., Tonry, J. L., Denneau, L. Flewelling, H., Stalder, B., Rest, A., Smith, K. W., Smartt, S. J., \& Weiland, H. 2018, \aj, 156, 241
\bibitem[Hensberge et al.(1977)]{Hensberge1977} Hensberge, E., De Loore, C., Zuiderwijk, E. J., \& Hammerschlag-Hensberge, G. 1977, \aap, 54, 443
  %Photometry of 'Silicon' stars HD 3580, HD 187473, HD 206653, HD 207188,
  % and HD 212432. All were found variable in the uvby bands, and
  % the intended comparison star HD 185183 was also found to be an
  % alpha2 CVn variable. Of these, HD 207188 shows a clear UCBH lightcurve,
  % but no other object does.
\bibitem[H\"{u}mmerich et al.(2018)]{Hummerich2018} H\"{u}mmerich, S., Niemczura, E., Walczak, P., Paunzen, E., Bernhard, K., Murphy, S. J., \& Drobek, D. 2018, \mnras, 474, 2467
  %Useful quote: "Most elements sink under the influence of gravity; however, those with
  %absorption lines near the local flux maximum are radiatively driven outward."
\bibitem[Kochukhov \& Wade(2010)]{Kochukhov2010} Kochukhov, O., \& Wade, G. A. 2010, \aap, 513, A13
% "Magnetic Doppler imaging of alpha^2 Canum Venaticorum in all four Stokes parameters"
  % Zeeman Doppler imaging of protoype alpha^2 CVn itself.
  % Found lower abundance of heavy elements at the negative magnetic pole
\bibitem[Kochukhov et al.(2015)]{Kochukhov2015} Kochukhov, O., Rusomarov, N., Valenti, J. A., Stempels, H. C., Snik, F., Rodenhuis, M., Piskunov, N., Makaganiuk, V., Keller, C. U., \& Johns-Krull, C. M. 2015, \aap, 574, A79
  %"Magnetic field topology and chemical spot distributions in the extreme Ap star HD 75049"
  %It is widely believed that magnetic field topologies of AP stars are predominately
  %axisymmetric and roughly dipolar, with the magnetic axis frequently misaligned relative
  %to the stellar rotation axis.
\bibitem[Lantz et al.(2004)]{SNIFS} Lantz, B., Aldering, G., Antilogus, P., et al. 2004, in Society of Photo-Optical Instrumentation Engineers (SPIE) Conference Series, Vol. 5249, Optical Design and Engineering, ed. L. Mazuray, P. J. Rogers, \& R. Wartmann, 146–155, doi: 10.1117/12.512493
%SNIFS instrument
\bibitem[Larson et al.(2003)]{CSS} Larson, S., Beshore, E., Hill, R., et al. 2003, DPS, 35, 3604
%Introductory presentation on the Catalina Sky Survey
\bibitem[Magnier et al.(2016a)]{PS1B} Magnier, E. A., Chambers, K. C., Flewelling, H. A., et al. 2016, arXiv:1612.05240
%PS1 Data Processing, Magnier, E. A., et al. 
\bibitem[Magnier et al.(2016b)]{PS1C} Magnier, E. A., Schlafly, E. F., Finkbeiner, D. P., et al. 2016, arXiv:1612.05242
%PS1 Calibration, Magnier, E. A., et al. 
\bibitem[Magnier et al.(2016c)]{PS1D} Magnier, E. A., Sweeney, W. E., Chambers, K. C., et al. 2016, arXiv:1612.05244
%PS1 Source Detection, Magnier, E. A., et al. 
\bibitem[Michaud (1970)]{Michaud1970} Michaud, Georges 1970, \apj, 160, 641
  %Groundbreaking suggestion that the peculiar abundances of Ap stars are due
  %to radiative levitation.
\bibitem[Michaud et al.(1981)]{Michaud1981} Michaud, G., M\'{e}gessier, C., \& Charland, Y. 1981, \aap, 103, 244
  %There are elements where radiative levitation actually exceeds the force of gravity.
  %Radiation pressure could potentially eject these from the star, but horizontal magnetic
  %fields could trap them well above the photosphere, where density is low and the lack
  %of collisions makes diffusion across magnetic field lines essentially impossible.
  %Individual atoms of elements that are usually ionized may spend a small fraction
  %of their time in a neutral state, which increases their ability to diffuse across
  %magnetic field lines, but apparently does not prevent trapping.
  %Other elements with less extreme radiative levitation would be concentrated where
  %field lines are vertical rather than horizontal. 
\bibitem[Ochsenbein et al.(2000)]{vizier} Ochsenbein, F., Bauer, P. \& Marcout, J. 2000, \apjs, 143, 23O
  %The vizier search engine.
\bibitem[Peterson (1970)]{Peterson1970} Peterson, D. M. 1970, \apj, 161, 685
  %The photometric variability of Ap stars. States that inhomogenous distribution
  %of Silicon *will* produce photometric variability by redirecting UV radiation
  %back into the visible -- hence, visible fluxes should be *positively correlated*
  %with the strength of the silicon lines. Mentions odd behavior of HD 221568.
\bibitem[Pickles (1998)]{Pickles1998} Pickles, A. J. 1998, \pasp, 110, 863
%A Stellar Spectral Flux Library: Pickles atlas of stellar spectra. 
\bibitem[Poretti et al.(1997)]{Poretti1997} Poretti, E., Koen, C., Martinez, P., Breuer, F., de Alwis, D., \& Haupt, H. 1997, \mnras, 292, 621
\bibitem[Preston (1974)]{Preston1974} Preston, G. W. 1974, \araa, 12, 257
%Foundationa review of chemically peculiar stars.
\bibitem[Press et al.(1992)]{nrc} Press, W. H., Teukolsky, S.A., Vetterling, W. T., \& Flannery, B. P. 1992, Numerical Recipes in C (Second Edition; New York, NY: Cambridge University Press)
  %NRC
\bibitem[Pyper(1969)]{Pyper1969} Pyper, D. M. 1969, \apjs, 164, 18
  %Amazing rotationally resolved abundance analysis for alpha2 CVn itself
\bibitem[Ryabchikova et al.(1990)]{Ryabchikova1990} Ryabchikova, T. A., Davidova, E. S., \& Adelman, S. J. 1990, \pasp, 102, 581
  %Spectrum variability of the Silicon Ap star HD 192913. Shows
  %variability in lines of Mg II, Si II, Ca II, Ti II, Cr II,
  %Mn II, Fe II, Sr II, and Eu II. Clear varations are seen in
  %all of these, but especially in Ti II, Mn II, and Cr II, and Eu II.
  %All of the spectral line variations seem to be in phase.
  %The star is also seen to vary in U-band photometry with an
  %amplitude of about 0.07 mag and a sawtooth lightcurve that
  %matches the waveform of the line strength variations, but
  %is neither strictly in phase with them nor strictly out of
  %phase with them. Rotation period is about 16.5 days.
  %Also discusses (and gives accurate wavelengths for) a fairly
  %large number of rare-Earth elements said to have been detected.
\bibitem[Ryabchikova et al.(2005)]{Ryabchikova2005} Ryabchikova, T., Leone, F., \& Kochukhov, O. 2005, \aap, 438, 973
  %Abundances and chemical stratification analysis in the atmosphere of
  %Cr-type Ap star HD 204411. Models chemical stratification and
  %high-SNR spectra at R=164,000 in daunting detail. Finds that
  %chemical *gradients* are *required* to match the spectra, and
  %obtains a much better match with such a model than with *any*
  %constant abundance that might be assumed. Counterinuitively,
  %finds decrements of most elements in the extreme upper atmosphere
  %but enhancements in the lower atmosphere at levels still shallow
  %enough to affect the spectra. Notes that the fastest-rotating
  %magnetic Ap stars are CU Vir (HD124424) and 56 Ari (HD 19832),
  %with periods of 0.52 days and 0.72 days, respectively.
\bibitem[Shappee et al.(2014)]{Shappee2014} Shappee, B. J., Prieto, J. L., Grupe, D., et al. 2014, \apj, 788, 48
  %First ASAS-SN paper, has nothing to do with variable stars.
\bibitem[Shulyak et al.(2010)]{Shulyak2010} Shulyak, D., Krti\v{c}ka, J., Mikul\`{a}\v{s}ek, Z., Kochukhov, O., \& L\"{u}ftinger, T. 2010, \aap, 524, A66
  %Successfully models ACVN lightcurves based on rotationally resolved
  %abundance measurements.
\bibitem[Sikora et al.(2019)]{Sikora2019} Sikora, J., Wade, G. A., Power, J., \& Neiner, C. 2019, \mnras, 483, 3127
%Volume-limited survey of mCP stars within 100 pc. Shows period
%distribution that is an excellent match to ATLAS UCBH stars.
\bibitem[Tonry et al.(2018a)]{Tonry2018} Tonry, J. L., Denneau, L., Heinze, A. N., Stalder, B., Smith, K. W., Smartt, S. J., Stubbs, C. W., Weiland, H. J., \& Rest, A. 2018, \pasp, 130, 4505
%ATLAS definition paper
\bibitem[Tonry et al.(2018b)]{refcat} Tonry, J. L., Denneau, L., Flewelling, H., Heinze, A. N., Onken, C. A., Smartt, S. J., Stalder, B., Weiland, H. J. \& Wolf, C. 2018, \apj, in press
%Refcat 2 paper.
%PS1 Pixel Processing, Waters, C. Z., et al. 
\bibitem[Wesemael et al.(1992)]{Wesemael1992} Wesemael, F., Fontaine, G., Bergeron, P., Lamontagne, R., \& Green, R. F. 1992, \aj, 104, 203
%Reference for hot subdwarf ATO J207.7199+36.7006 = PG 1348+369.

\end{thebibliography}
\end{document}